\documentclass[prl,twocolumn,superscriptaddress,float,aps]{revtex4-2}
\usepackage{bm,color,amsmath,amssymb,mathrsfs,latexsym,graphicx,psfrag,tabularx}
\usepackage[hidelinks,colorlinks,linkcolor=blue,citecolor=blue,urlcolor=blue]{hyperref}
\usepackage{etoolbox}
\makeatletter
\patchcmd{\@maketitle}{\@author}{\@author\show\@thanks}{}{}
\makeatother
\usepackage{dsfont}
\usepackage{xcolor}
\usepackage{enumitem}

\newcommand{\ii}{\mathrm{i}}
\newcommand{\dd}{\mathrm{d}}
\newcommand{\kk}{\mathbf{k}}

\newcommand{\tr}{\operatorname{tr}}
\newcommand{\Lmax}{N_{\max}}

\newcommand{\bra}[1]{\langle #1|}
\newcommand{\ket}[1]{|#1\rangle}
\newcommand{\braket}[2]{\langle #1|#2\rangle}

\newcommand{\ELLminus}{E^{\uparrow}_{-}}
\newcommand{\ELLplus}{E^{\uparrow}_{+}}
\newcommand{\ELLTRminus}{E^{\downarrow}_{-}}
\newcommand{\ELLTRplus}{E^{\downarrow}_{+}}
\newcommand{\ELLzero}{E^{\uparrow}_{0}}
\newcommand{\ELLTRzero}{E^{\downarrow}_{0}}
\newcommand{\ELLpm}{E^{\uparrow}_{\pm}}
\newcommand{\ELLTRpm}{E^{\downarrow}_{\pm}}

\newcommand{\meff}{m^{*}_{\mathrm{eff}}}
\newcommand{\Bstar}{B^{*}}
\newcommand{\dmu}{\delta\mu}

\newcommand{\Berry}{\Omega}
\newcommand{\metric}{g}

\newcommand{\kB}{k_{\!B}}

\newcommand{\cref}[1]{ref.~\ref{#1}}
\newcommand{\Cref}[1]{Ref.~\ref{#1}}

\usepackage{mathtools}
\usepackage{amsthm}

\providecommand{\mel}[3]{\langle #1|#2|#3\rangle}

\begin{document}

\title{Lifshitz--Kosevich Theory of Anomalous Landau Levels in Topological Flat Bands}
\author{Chao-Xing Liu}
\affiliation{%
	Department of Physics, The Pennsylvania State University, University Park, Pennsylvania 16802, USA
}
\affiliation{Center for Theory of Emergent Quantum Matter, The Pennsylvania State University, University Park, Pennsylvania 16802, USA}
\date{\today}

\begin{abstract}
	In conventional metals, quantum oscillations arise from Landau quantization of Fermi-surface cyclotron orbits, whose dynamics are governed by the Fermi velocity and cyclotron effective mass within Lifshitz--Kosevich (LK) theory. A perfectly flat band, by contrast, has vanishing group velocity, which would na\"ively imply an infinite cyclotron mass and complete thermal suppression of quantum oscillations. Yet topological flat bands can support anomalous Landau levels (LLs) whose finite-field spacing is generated by quantum geometry rather than band curvature, allowing quantum oscillations to persist. This work addresses how such anomalous flat-band LLs behave within the LK framework and whether their thermal damping can reveal quantum-geometric information. Using a minimal model with exactly flat topological bands, we derive an LK theory for these anomalous LLs and analyze fixed-density magnetization oscillations. The resulting oscillations exhibit a finite LK effective mass that is substantially larger than the normal-band value and possesses a strong magnetic-field dependence. In the weak-field limit, this anomalous mass reflects the quantum-geometric origin of the LL spacing and scales inversely with both the magnetic field and the trace of the quantum metric. Thus, thermal damping of flat-band quantum oscillations directly measures the quantum metric, establishing quantum oscillations as a probe to flat-band quantum geometry.
\end{abstract}

\maketitle


{\it Introduction - }
Quantum oscillations are conventionally understood as a Fermi-surface phenomenon~\cite{shoenberg1984magnetic}: in a magnetic field, cyclotron orbits are quantized into Landau levels, and their passage through the Fermi energy produces oscillatory thermodynamic and transport responses. 
The standard Lifshitz--Kosevich (LK) formalism~\cite{lifshitz1956theory,shoenberg1984magnetic} expresses
the oscillatory part of the grand potential $\delta \Omega$ at a finite temperature as a sum over harmonics,
\begin{equation}
	\delta \Omega \propto \sum_{p=1}^{\infty}
	\frac{R_T(p)}{p^{2}}
	\cos (2\pi p\,F/B - \phi),
	\label{eq:lk-standard}
\end{equation}
where $p$ is the harmonic index, $F$ and $\phi$ define the oscillation frequency and phase, respectively, and $B$ is external magnetic field. The thermal damping factor $R_T(p) = X_p/\sinh X_p$ depends on
$X_p = 2\pi^{2} p \kB T / \hbar\omega_{c}$, with the cyclotron energy
$\hbar\omega_{c} = e\hbar B / m^{*}$ set by the cyclotron mass $m^{*}$, which
therefore fixes the characteristic temperature scale of thermal smearing. In
systems with vanishing band dispersion, however, the band mass diverges.
Applied literally to a perfectly flat band, $m^{*}\!\to\!\infty$ gives
$X_p\!\to\!\infty$ and $R_T\!\to\!0$, so the conventional LK formula predicts
complete thermal suppression and a total absence of quantum oscillations. This conclusion, however, rests on identifying the thermal damping scale with a
cyclotron energy set by the band dispersion---an identification that need not hold
for topological flat bands, which have recently been realized in moir\'e materials~\cite{bistritzer2011moire,cao2018unconventional,serlin2020intrinsic,song2022magic,ledwith2020fractional,cai2023signatures,zeng2023thermodynamic,xu2023observation,adak2024tunable,mak2022semiconductor,balents2020superconductivity,liu2021orbital,checkelsky2024flat,torma2022superconductivity,song2019all,song2021twisted,po2019faithful,ahn2019failure}. Whether quantum oscillations persist in such a band,
and how the LK damping factor should be reformulated when the conventional dispersive cyclotron energy is absent, remain unresolved.

Recently, it was recognized that the Landau levels of a perfectly flat band under magnetic fields may not remain degenerate but instead can spread over a finite energy window. The magnitude of this anomalous Landau-level spreading is set by the quantum geometry of the Bloch states~\cite{kang2025measurements,kim2025direct,yu2025quantum,torma2023essay,liu2025quantum} ---quantified by the Hilbert--Schmidt quantum distance and the cross-gap Berry connection---rather than by any band dispersion~\cite{rhim2020quantum,hwang2021geometric,rhim2020singular,long2026interplay,jung2024quantum,kawakami2025singular,datta2025anomalous,fu2026anomalous}.
This spreading endows even a perfectly flat band with a finite, field-dependent energy scale that has no counterpart in the dispersive cyclotron picture, reopening the possibility that quantum oscillations survive in the flat-band limit---now controlled by quantum geometry rather than by band curvature.
This raises two questions for topological flat bands: what replaces the
cyclotron energy in the LK thermal damping of anomalous LL oscillations, and
can that damping be used to extract quantum-geometric information?

To address these questions, we develop an LK theory for quantum oscillations of anomalous LLs in a minimal model with perfectly flat topological bands~\cite{liu2025ideal}. Our central claim is that the LK thermal damping for anomalous LL oscillations is controlled by the \emph{local} LL spacing at the Fermi level, $v_\mu^{\alpha,\eta}=\bigl|\partial E^\alpha_\eta(n,B)/\partial n\bigr|_{n^*_{\alpha,\eta}}$, where $E^\alpha_\eta$ is the branch-resolved Landau-level energy and $n$ is the Landau-level index treated as a continuous variable. We compare fixed-density magnetization oscillations of normal LLs, associated with dispersive bands, with anomalous LLs from topological flat bands. We compute their temperature dependence, extract the thermal damping from local oscillation windows, and express the results in the conventional effective-mass framework. Although the anomalous LL oscillations exhibit much stronger thermal damping, the fitted effective masses remain finite, are approximately an order of magnitude larger than the normal-band masses in the representative comparison, and exhibit a clear monotonic field dependence, in contrast with the nearly constant normal-band masses. Expanding the exact LL spacing in the weak-field limit yields $v_\mu^{\alpha,-} \sim B^2 \tr\metric$, so that $\meff \sim 1/(B\,\tr\metric)$, where $B$ is the magnetic field and $\metric$ is the quantum metric. The LK thermal scale of anomalous oscillations is thus directly tied to the quantum metric of the topological flat band. Taken together, these results answer both questions: \emph{(i)} anomalous flat-band oscillations persist within the LK framework, with the local LL spacing $v_\mu^{\alpha,\eta}$ replacing the cyclotron energy and yielding a finite, strongly field-dependent effective mass larger than the dispersive-band value; and \emph{(ii)} their thermal damping provides a probe of flat-band quantum geometry.

\begin{figure}[t]
	\centering
	\includegraphics[width=\columnwidth]{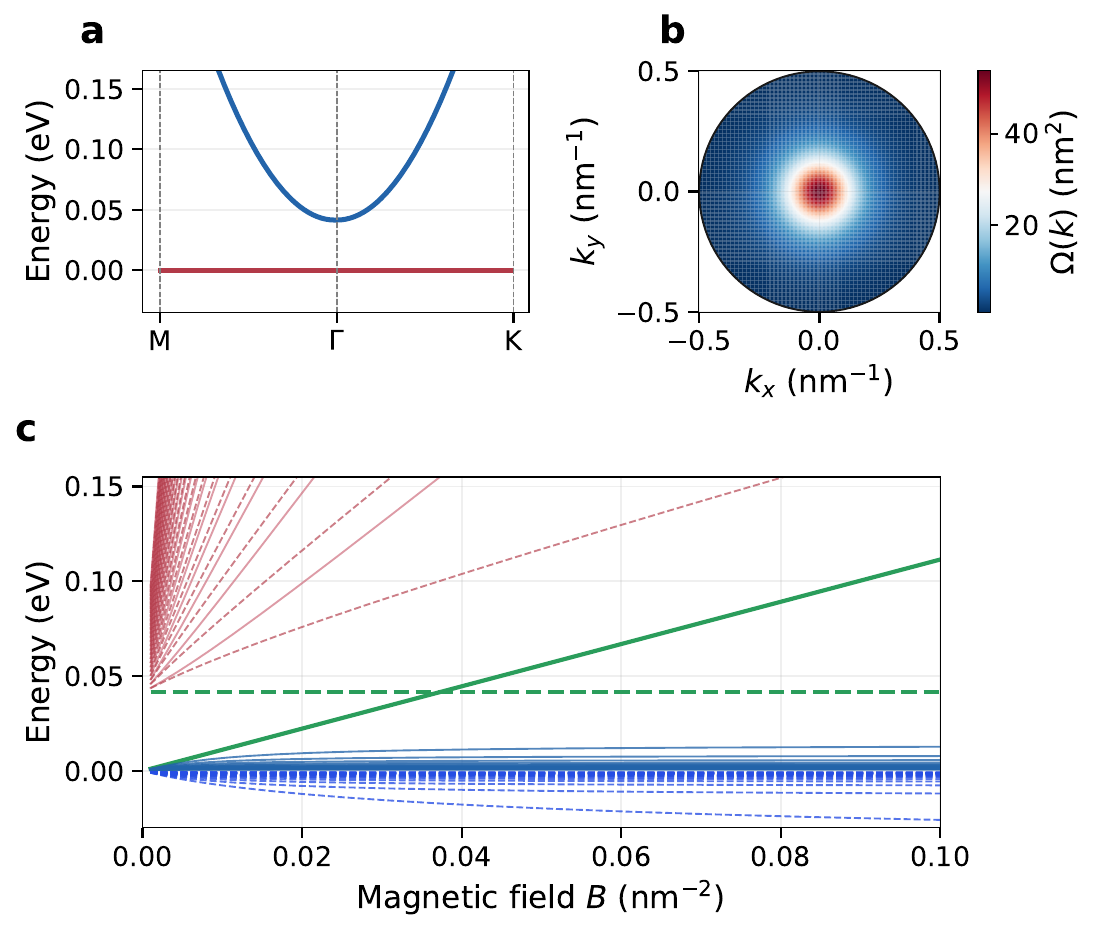}
	\caption{(a) Band dispersion at the ideal flat-band point along the $M$--$\Gamma$--$K$ path; the dispersive and flat bands are blue and red, respectively. (b) Berry-curvature distribution of the spin-up flat band. (c) Landau-level spectrum; normal, anomalous, and zeroth LLs are red, blue, and green, respectively, while solid and dashed curves distinguish the two spin sectors. Parameters: $a=1.115\,\mathrm{eV\cdot nm^2}$, $c=0.215\,\mathrm{eV\cdot nm}$, $E_0=0\,\mathrm{eV}$.}
	\label{fig:main1}
\end{figure}

{\it Ideal topological flat bands - } We start with a minimal model with perfectly flat topological bands, which can describe low-energy physics of moir\'e heterostructures with type-II band alignment in Ref.~\cite{liu2025ideal}. This model consists of heavy (flat) fermion and light (dispersive) fermion components, 
and the corresponding Hamiltonian is given by 
\begin{equation}
	H_\alpha(\kk) =
	\begin{pmatrix}
		E_0 + a k^{2} & \alpha c k_{\alpha} \\
		\alpha c k_{-\alpha}    & E_0 + \Delta_f
	\end{pmatrix},
	\label{eq:thf-hamiltonian}
\end{equation}
where $\alpha=\uparrow, \downarrow$ or $\pm$ labels two opposite spin states, $k_{\pm} = k_x \pm \ii k_y$, $a$ defines the band-curvature coefficient of the dispersive bands, $c$ is the coupling between the flat and dispersive bands, and $E_{0}$ is the energy reference and $\Delta_f$ is the energy separation between the flat and dispersive bands at $\kk=0$. The above Hamiltonian is valid within the momentum cutoff $\Lambda$, e.g. $|\kk| \le \Lambda$, which is set by the moir\'e Brillouin zone size in moir\'e heterostructures. In our numerical results, energies and the thermal scale $k_B T$ are measured in $\mathrm{eV}$ and lengths in $\mathrm{nm}$. 
Two spin Hamiltonians $H_\uparrow$ and $H_\downarrow$ are related by time-reversal (TR) symmetry, e.g. $H_\downarrow(\kk) = H_\uparrow^*(-\kk)$, so the full system satisfies TR. When the condition $\Delta_f = \frac{c^{2}}{a}$ is satisfied, the lower band is perfectly flat with $E^{\uparrow (\downarrow)}_{-}(\kk) = E_0$ and the upper band is dispersive with a finite gap, given by $E^{\uparrow (\downarrow)}_{+}(\kk) = E_0 + \frac{c^{2}}{a} + a k^{2}$, as shown in Fig.~\ref{fig:main1}a. The full effective Hamiltonian possesses both TR and inversion symmetries, so the eigen-states for spin up and down Hamiltonian are degenerate at each momentum. For the perfectly flat lower band $E^{\uparrow (\downarrow)}_-$, its wavefunction carries nontrivial quantum geometry, as demonstrated by the Berry-curvature distribution in Fig.~\ref{fig:main1}b. In particular, the lower band satisfies the ideal quantum geometry condition
\begin{align}
	\tr\metric_{\alpha}(\kk) = |\Berry_{\alpha}(\kk)|
	= \frac{2c^{2}}{(a k^{2} + c^{2}/a)^{2}}.
	\label{eq:geometry-results}
\end{align}
Thus, the two TR-related sectors have the same quantum metric and opposite Berry curvature; the spin-up (spin-down) flat band has Chern number $+1$ ($-1$).
More discussion on the quantum geometry of ideal topological flat bands can be found in Sec.A of Supplementary Materials (SM). 




We next consider this minimal model under a perpendicular magnetic field $\mathcal{B}$, which enters through minimal coupling $\kk\to\boldsymbol{\Pi} = \kk + e\mathbf{A}/\hbar$, where $\mathbf{A} = \frac{\mathcal{B}}{2}(-y,x,0)$ is the vector potential for the symmetric gauge.
The canonical momentum operators satisfy $[\Pi_x,\Pi_y]=-\ii B$ with $B = e\mathcal{B}/\hbar$, so the ladder operators can be defined as $b = \Pi_-/\sqrt{2B}$, $b^{\dagger} = \Pi_+/\sqrt{2B}$ with $[b,b^{\dagger}]=1$. Here, \(B\) is measured in \(\mathrm{nm}^{-2}\). Using these operators, we solve the Hamiltonian in Eq.~\eqref{eq:thf-hamiltonian} (see SM Sec.~B). The spectrum in Fig.~\ref{fig:main1}(c) comprises two zeroth LLs,
\begin{equation}
	\ELLzero(B) = E_0 + aB,\qquad
	\ELLTRzero(B) = E_0 + \frac{c^{2}}{a} = E_0 + \Delta_f,
	\label{eq:zero-modes}
\end{equation}
and four LL branches for $n\ge 1$, given by
\begin{equation}
	\ELLpm(n,B) = E_0 + \tfrac12\!\left[U_n \pm \sqrt{U_n^{2} - 4c^{2}B}\right],
	\label{eq:ELL-energies}
\end{equation}
with $U_n \equiv c^{2}/a + (2n+1)aB$, and
\begin{equation}
	\ELLTRpm(n,B) = E_0 + \tfrac12\!\left[V_n \pm \sqrt{V_n^{2} + 4c^{2}B}\right],
	\label{eq:ELLTR-energies}
\end{equation}
with $V_n \equiv c^{2}/a + (2n-1)aB$.
We note that the zero mode $\ELLzero$ linearly increases with $B$, while $\ELLTRzero$ is independent of $B$. Moreover, the two lower-energy branches ($\ELLminus$, $\ELLTRminus$) form the anomalous LL manifold, originating from topological flat-bands as $E^{\uparrow,\downarrow}_{-}(B\rightarrow 0) = E_0$, while the two higher-energy branches ($\ELLplus$, $\ELLTRplus$) are the normal LLs from dispersive bands. The gaps between adjacent anomalous LLs, defined as $\Delta E^{\uparrow (\downarrow)}_n = E^{\uparrow (\downarrow)}_-(n+1)- E^{\uparrow (\downarrow)}_-(n)$, scales as $\Delta E^{\uparrow (\downarrow)}_n\propto B^{2}$ for a small $B$, slower than the linear-in-$B$ growth of the normal LLs from dispersive bands. For a large $n$, we can treat $n$ as a continuous variable to express the Fermi-level local LL spacing as
\begin{eqnarray}
   v_\varepsilon^{\alpha,\eta}
   =
   \left|\frac{\partial E^\alpha_\eta(n,B)}{\partial n}\right|_{n(\varepsilon)},
   \qquad
   E^\alpha_\eta(n,B)=\varepsilon,
   \label{eq:local-LL-spacing}
\end{eqnarray}
which plays a crucial role in the thermal damping of the LK theory described below. 


%

{\it Quantum oscillations of anomalous LLs - } We apply the standard LK
theory~\cite{lifshitz1956theory,shoenberg1984magnetic} to our model with both
normal and anomalous LLs (see SM Sec.~C) and derive the oscillatory part of the
grand potential as
\begin{equation}
\delta \Omega = \frac{D}{2\pi^{2}}\sum_{\alpha, \eta}\sum_{p=1}^{\infty}\frac{v_\mu^{\alpha,\eta}}{p^{2}} R_T \cos \bigl[2\pi p n^*_{\alpha,\eta}\bigr], \label{eq:omega-tilde}
\end{equation}
where $D = B/2\pi$ is the LL degeneracy, the LK thermal damping factor is given by
\begin{equation}
	R_T(p) = \frac{X_p}{\sinh X_p},\qquad
	X_p = \frac{2\pi^{2} p\, k_B T}{v_\mu^{\alpha,\eta}},
	\label{eq:Xp-manuscript}
\end{equation}
and the LL filling $n^*_{\alpha,\eta}$ is fixed by the chemical potential
$\mu$ through $E^\alpha_\eta(n^*_{\alpha,\eta},B) = \mu$.
Equations~(\ref{eq:omega-tilde}) and (\ref{eq:Xp-manuscript}) show that the
thermal damping is controlled by the local spacing $v_{\mu}^{\alpha,\eta}$ at
the chemical potential. In our numerical calculations, we fix the carrier
density $\rho$, which determines $\mu$ and $n^*_{\alpha,\eta}$ at each
magnetic field $B=e\mathcal{B}/\hbar$. The magnetization is evaluated as
$M=-(e/\hbar)\,\partial\Omega/\partial B$ at fixed $\rho$.

\begin{figure}[t]
	\centering
	\includegraphics[width=\columnwidth]{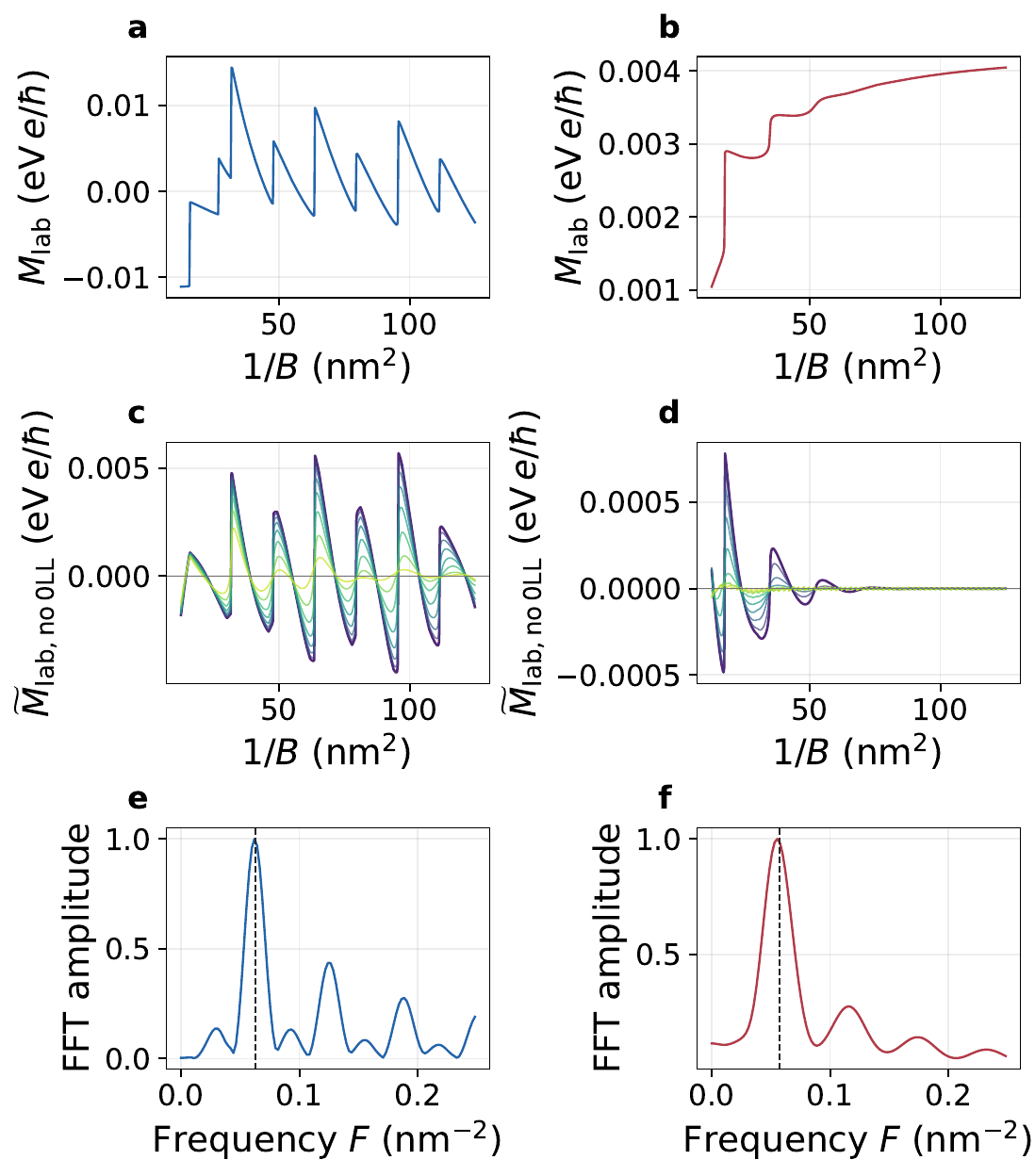}
	\caption{Fixed-density magnetization oscillations. (a) Full physical-field magnetization $M_{\mathcal{B}}$ in the normal regime, $\rho=+0.01\,\mathrm{nm}^{-2}$. (b) Full physical-field magnetization in the anomalous regime, $\rho=-0.15\,\mathrm{nm}^{-2}$. (c,d) Corresponding background-subtracted oscillatory magnetization at $k_B T=(2.00,\,3.07,\,4.71,\,7.22,\,11.1,\,17.0,\,26.1,\,40.0)\times10^{-4}\,\mathrm{eV}$, ordered from purple to yellow. (e,f) Corresponding normalized FFT spectra; dashed vertical lines mark the active-density counting frequency $F_{\rho}=2\pi|\rho_{\mathcal S}|$, measured in $\mathrm{nm}^{-2}$. Blue and red denote the normal and anomalous regimes, respectively.}
	\label{fig:main2}
\end{figure}

Figures~\ref{fig:main2}(a) and \ref{fig:main2}(b) show the magnetization
$M_{\mathcal{B}}$ as a function of $1/B$ for normal LLs in the electron-doped
regime, $\rho=+0.01\,\mathrm{nm}^{-2}$, and anomalous LLs in the hole-doped
regime, $\rho=-0.15\,\mathrm{nm}^{-2}$, respectively. Subtracting a smooth
background isolates the oscillatory magnetization in
Figs.~\ref{fig:main2}(c) and \ref{fig:main2}(d). The anomalous LLs retain
visible oscillations at high field, but their amplitude decays much more
rapidly than that of the normal LLs as $B$ decreases. The FFT peaks in
Figs.~\ref{fig:main2}(e) and \ref{fig:main2}(f) agree with the fundamental
counting frequency $F_{\rho}=2\pi|\rho_{\mathcal S}|$, where
$\rho_{\mathcal S}$ is the active-carrier density in the electron,
shallow-hole, or deep-hole sector (see SM Sec.~D). Although a Fermi-surface
area is not defined for the flat band, its oscillation frequency still follows
from LL counting: each filled LL accommodates $D=B/(2\pi)$ carriers per unit
area, so $n^*\simeq2\pi|\rho_{\mathcal S}|/B$. The agreement between the FFT
peaks and the dashed counting references therefore does not depend on a band
dispersion. Additional frequency diagnostics are presented in SM Sec.~D.

%
%

\begin{figure}[t]
	\centering
	\includegraphics[width=\columnwidth]{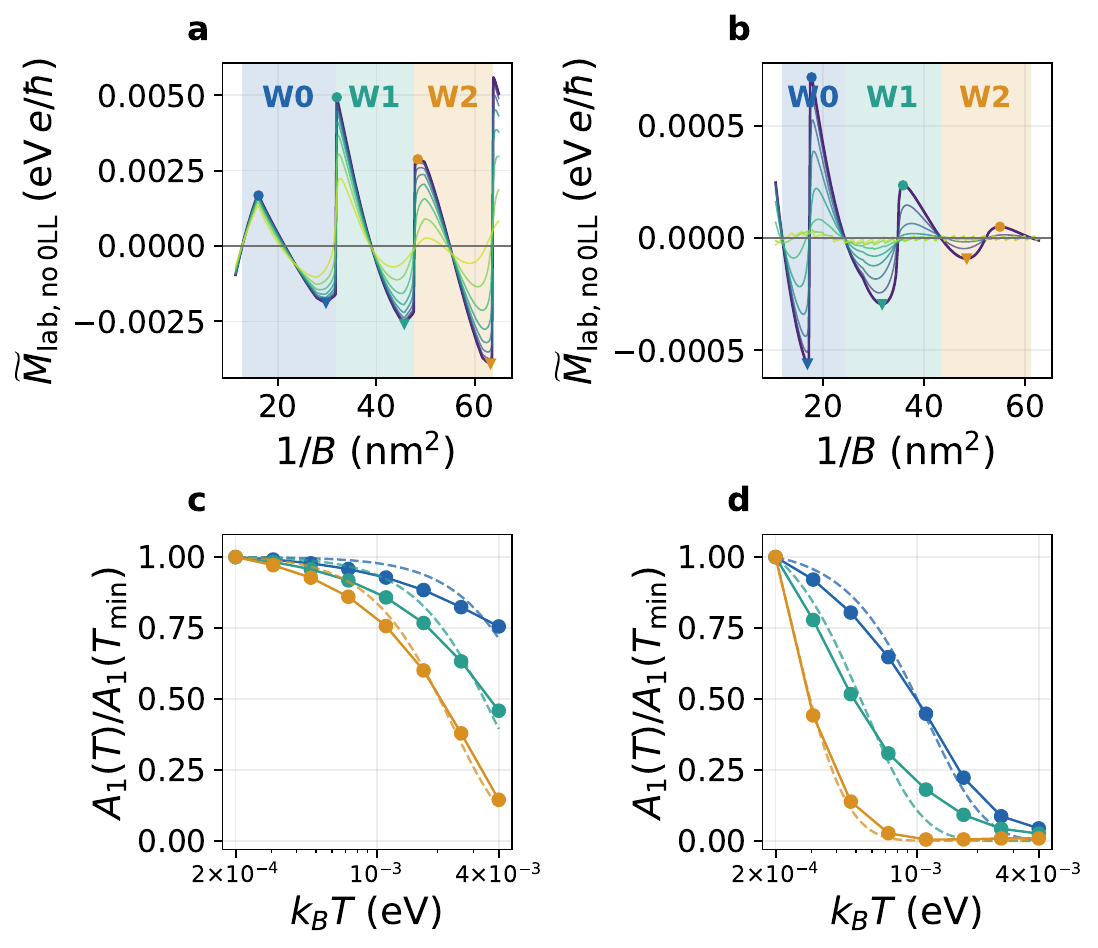}
	\caption{Thermal damping extracted from local oscillation windows. (a,b) Oscillatory magnetization in the normal regime, $\rho=+0.01\,\mathrm{nm}^{-2}$, and anomalous regime, $\rho=-0.15\,\mathrm{nm}^{-2}$, respectively, at $k_B T=(2.00,\,3.07,\,4.71,\,7.22,\,11.1,\,17.0,\,26.1,\,40.0)\times10^{-4}\,\mathrm{eV}$, ordered from purple to yellow. In (a,b), $W_0$, $W_1$, and $W_2$ are blue, teal, and orange, respectively. (c,d) Normalized $p=1$ harmonic amplitudes for the windows in (a,b), respectively; dashed curves are standard-LK fits.}
	\label{fig:main3}
\end{figure}

\begin{figure}[t]
	\centering
	\includegraphics[width=\columnwidth]{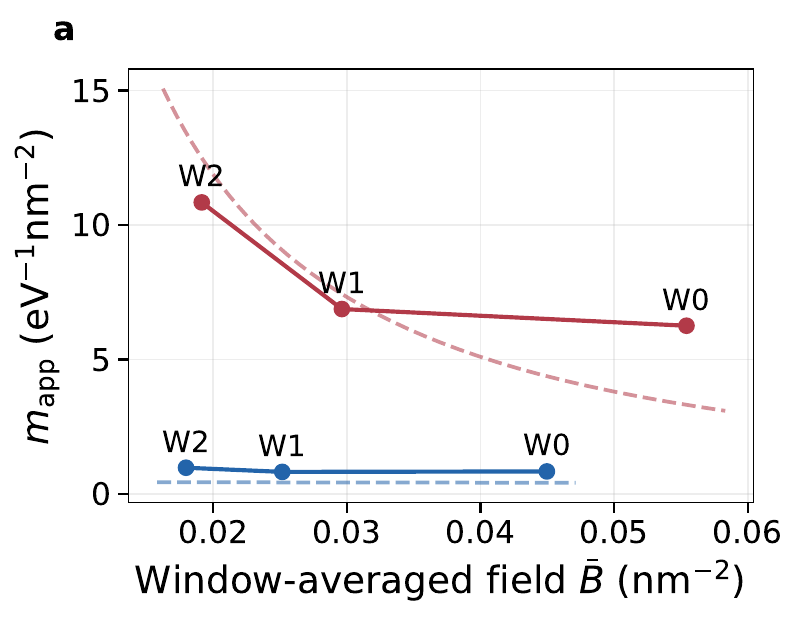}
	\caption{(a) Apparent effective mass, measured in $\mathrm{eV}^{-1}\mathrm{nm}^{-2}$, versus window-averaged magnetic field. Solid markers show masses extracted from the Fig.~\ref{fig:main3} damping fits for the normal regime, $\rho=+0.01\,\mathrm{nm}^{-2}$ (blue), and anomalous regime, $\rho=-0.15\,\mathrm{nm}^{-2}$ (red). Dashed curves show the corresponding analytical LK trends; labels identify the local windows.}
	\label{fig:main4}
\end{figure}

{\it Thermal damping and effective mass - } Finite temperature suppresses the
magnetization oscillations in Figs.~\ref{fig:main2}(c) and
\ref{fig:main2}(d). Because the signals are not perfectly sinusoidal, a global
Fourier transform does not cleanly isolate the thermal damping of individual
harmonics. We therefore use the short $1/B$ windows $W_0$, $W_1$, and $W_2$
marked in Figs.~\ref{fig:main3}(a) and \ref{fig:main3}(b), corresponding to
the first three complete high-field oscillation periods for the normal and
anomalous LLs. Within each window, we extract the amplitude $A_1$ of the
fundamental $p=1$ harmonic and plot $A_1(T)/A_1(T_{\min})$ in
Figs.~\ref{fig:main3}(c) and \ref{fig:main3}(d). The anomalous oscillations
show substantially stronger thermal damping. We fit this suppression with the
standard LK form using a single effective mass $\meff$:
\begin{equation}
	\frac{A_1(T)}{A_1(T_{\min})}
	= \frac{R_T(T,\meff,\bar{B}_w)}{R_T(T_{\min},\meff,\bar{B}_w)},\qquad
	X_1 = \frac{2\pi^{2} \kB T \meff}{\bar{B}_w},
	\label{eq:lk-fit}
\end{equation}
where $\bar{B}_w$ is the average magnetic field in the window. The dashed
curves in Figs.~\ref{fig:main3}(c) and \ref{fig:main3}(d) show that this
one-parameter form captures the qualitative damping trend in both regimes.
Comparing Eq.~(\ref{eq:lk-fit}) with Eq.~(\ref{eq:Xp-manuscript}) gives
\begin{equation}
	\meff(\bar{B}_w,\mu) = \frac{\bar{B}_w}{v_\mu^{\alpha,\eta}}, 
	\label{eq:meff-def}
\end{equation}
in units of $\mathrm{eV}^{-1}\mathrm{nm}^{-2}$. Figure~\ref{fig:main4}
summarizes the extracted masses. For the normal regime
($\rho=+0.01\,\mathrm{nm}^{-2}$), the fitted values are
$\meff\approx(0.626,\,0.788,\,0.958)\,\mathrm{eV}^{-1}\mathrm{nm}^{-2}$
across $W_0,W_1,W_2$ and remain comparatively small. In the anomalous regime
($\rho=-0.15\,\mathrm{nm}^{-2}$), they increase to
$\meff\approx(5.92,\,6.73,\,11.0)\,\mathrm{eV}^{-1}\mathrm{nm}^{-2}$ and
show a clear monotonic trend across the windows. The contrasting magnitude and
field dependence provide an experimental signature distinguishing anomalous
from normal LL oscillations.

The different trends can be understood from Fig.~\ref{fig:main1}(c). At large
$n^*$, the normal branches satisfy $(\mu-E_0)^2\gg c^2B$, whereas the
anomalous branches satisfy $(\mu-E_0)^2\ll c^2B$. The normal effective mass
then approaches the band-curvature value $\meff\approx1/(2a)$. For the
anomalous branches,
$\meff\approx c^2B/[2a(\mu-E_0)^2]$. At fixed carrier density, $n^*$ is
approximately fixed and $\mu-E_0\propto B$ in the anomalous regime, yielding
$\meff\sim1/B$. The dashed curves in Fig.~\ref{fig:main4} show these
analytical trends and agree qualitatively with the numerical results; the
derivation and additional tests are given in SM Secs.~C and D.

As the spreading of anomalous LLs is directly connected to quantum geometry of topological flat bands, we finally explore how to connect the effective mass $\meff$ to quantum geometric information. In the semi-classical limit with a large LL filling $n^*_{\alpha,-}\gg 1$, we find $U_{n^*} \approx V_{n^*} \approx a k_{n^*}^{2} + c^{2}/a$ with $k_{n^*}^{2} \approx 2n^*_{\alpha,-} B$. Consequently, we find the anomalous-branch local LL spacing $v_\mu^{\alpha,-} \approx \frac{2a c^{2} B^{2}}{(c^{2}/a + a k_{n^*}^{2})^{2}} = a B^{2} \tr\metric_\alpha(k_{n^*}) $ for both $\alpha = \uparrow, \downarrow$. This establishes the relationship between the effective cyclotron mass defined in Eq.(\ref{eq:meff-def}) and the quantum metric, 
\begin{equation}
	\meff \approx \frac{1}{a B\,\tr\metric_\alpha(k_{n^*})}. 
	\label{eq:meff-geometric}
\end{equation} 
Therefore, we demonstrate the magnetic field dependence of the effective mass for anomalous LLs is directly tied to the quantum metric of topological flat bands. 

{\it Conclusion - } We have developed an LK description of quantum
oscillations from the anomalous Landau levels of ideal topological flat bands.
The oscillations remain finite because their thermal scale is set by the local
LL spacing rather than by a dispersion-derived cyclotron energy. In the
representative fixed-density comparison, the anomalous oscillations damp much
more rapidly than the normal oscillations and yield fitted effective masses
approximately an order of magnitude larger, with a pronounced field
dependence. In the semiclassical large-$n^*$ limit,
$\meff\sim1/[B\,\tr\metric]$, directly relating the damping scale to the
quantum metric. The frequency and the thermal scale thus play complementary
roles: LL counting fixes the oscillation period, while quantum geometry governs
the anomalous thermal damping. Given recent experimental progress in moir\'e
materials~\cite{cao2018unconventional,serlin2020intrinsic,cai2023signatures,zeng2023thermodynamic,xu2023observation,liu2022quantum},
temperature-dependent quantum-oscillation measurements offer a concrete route
to probing the quantum geometry of topological flat bands.

{\it Data and code availability - }
The code, data, and reproduction instructions associated with this work are
available in the Agentic Publication Protocol (APP) repository at
\url{https://github.com/lcxlight/lk-anomalous-landau-levels}.

{\it Acknowledgments - }
We thank Zhen Bi and Jainendra Jain for helpful discussions. The author also acknowledges the use of Claude Code 2.1.200, OpenAI Codex CLI 0.140.0-alpha.2/GPT-5-based Codex, and TEXRA 0.39.2, directed through author-written prompts and author-provided manuscript/source files, for assistance with manuscript editing, code organization, numerical-code development, analytical derivation checks, reproducibility checks, and bibliography management; all AI-assisted code, calculations, derivations, references, scientific conclusions, and final manuscript content were independently checked by and remain the responsibility of the author. C.-X.L. acknowledges support from NSF grant DMR-2241327 and the support from ICDS Faculty Upskilling award ICDS\_UP25\_028016 from Penn State's Institute for Computational and Data Sciences (RRID:SCR\_025154)


\clearpage
\onecolumngrid
\appendix

\setcounter{secnumdepth}{3}
\renewcommand{\thesection}{\Alph{section}}
\renewcommand{\thesubsection}{\thesection.\arabic{subsection}}
\numberwithin{equation}{section}
\numberwithin{figure}{section}


\section{Model and Quantum Geometry Details}
\label{app:model}

\subsection{Flat-band condition and the ideal point}

We consider the eigen-equation 
\begin{eqnarray}
    H_{\alpha}(\kk)\ket{u^\alpha_{\pm}(\kk)} = E^\alpha_{\pm}(\kk)\ket{u^\alpha_{\pm}(\kk)},
    \label{eq:eigen-eq-app}
\end{eqnarray}
and work with the shifted Hamiltonian
\begin{equation}
    \widetilde{H}_\alpha(\kk) = H_{\alpha}(\kk) - E_0\,\mathbb{I}
    =
    \begin{pmatrix}
        a k^{2} & \alpha c k_{\alpha} \\
        \alpha c k_{-\alpha} & \Delta_f
    \end{pmatrix},
    \qquad
    \Delta_f \equiv E_1 - E_0,
    \label{eq:h-tilde-app}
\end{equation}
where $k_{\pm} = k_x \pm \ii k_y$, $k^2 = k_x^2+k_y^2$, and the two spin sectors $\alpha = \uparrow,\downarrow$ (or equivalently $\alpha=\pm 1$) are related by time-reversal.
Starting from Eq.~\eqref{eq:h-tilde-app}, the characteristic polynomial is
\begin{align}
    \det[\widetilde H_\alpha(\kk) - \lambda \mathbb{I}]
     & = (a k^{2} - \lambda)(\Delta_f - \lambda) - c^{2} k_{+} k_{-}
    \nonumber                                                                  \\
     & = \lambda^{2} - (a k^{2} + \Delta_f)\lambda + (a\Delta_f - c^{2})k^{2}.
    \label{eq:charpoly-app}
\end{align}
For a flat band, $\lambda\equiv0$ for all $\kk$.
Substituting $\lambda=0$ gives $(a\Delta_f - c^{2})k^{2}=0$, which requires $\Delta_f = c^{2}/a$, i.e.\ $E_1 = E_0 + c^{2}/a$.
With this condition, Eq.~\eqref{eq:charpoly-app} factorises as $\lambda[\lambda-(a k^{2}+c^{2}/a)]=0$, yielding the band energies
\begin{equation}
    E^\alpha_{-}(\kk) = E_0,\qquad E^\alpha_{+}(\kk) = E_0 + a k^{2} + \frac{c^{2}}{a}.
    \label{eq:band-energies-app}
\end{equation}

\paragraph{Bloch states.}
Define $N(\kk) \equiv a^{2}k^{2} + c^{2}$, and the eigen-wavefunctions are given by
\begin{equation}
    \ket{u^\alpha_{-}(\kk)}
    =
    \frac{1}{\sqrt{N(\kk)}}
    \begin{pmatrix}
        -c \\ \alpha a k_{-\alpha}
    \end{pmatrix},
    \qquad
    \ket{u^\alpha_{+}(\kk)}
    =
    \frac{1}{\sqrt{N(\kk)}}
    \begin{pmatrix}
        \alpha a k_{\alpha} \\ c
    \end{pmatrix},
    \label{eq:bloch-states-app}
\end{equation}
One verifies the orthonormal condition, e.g. $\langle u^\alpha_{\pm}|u^\alpha_{\pm}\rangle = 1$ and $\langle u^\alpha_{+}|u^\alpha_{-}\rangle = 0$.

\subsection{Quantum geometric tensor}

For the isolated flat band in sector $\alpha$, the projector onto the complementary band is $1-\ket{u^\alpha_{-}}\bra{u^\alpha_{-}} = \ket{u^\alpha_{+}}\bra{u^\alpha_{+}}$.
Hence, one can define the quantum geometry tensor as
\begin{equation}
    Q^\alpha_{ij}(\kk)
    = \braket{\partial_{i}u^\alpha_{-}}{u^\alpha_{+}} \braket{u^\alpha_{+}}{\partial_{j}u^\alpha_{-}}.
    \label{eq:QGT-app}
\end{equation}
The matrix elements are computed via the Hellmann--Feynman (HF) relation.
Define the shifted interband energy
\begin{equation}
    \Delta(k) \equiv E^\alpha_{+}(\kk)-E_0 = a k^2+\frac{c^2}{a}.
    \label{eq:Delta-app}
\end{equation}
Since $\widetilde{H}_\alpha(\kk)\ket{u^\alpha_{-}} = 0$, differentiating with respect to $k_i$ gives
$(\partial_{i}\widetilde{H}_\alpha)\ket{u^\alpha_{-}} + \widetilde{H}_\alpha\ket{\partial_{i}u^\alpha_{-}} = 0$.
Taking $\bra{u^\alpha_{+}}$ from the left and using $\bra{u^\alpha_{+}}\widetilde{H}_\alpha = \bra{u^\alpha_{+}}(E^\alpha_{+}-E_{0}) = \Delta(k)\bra{u^\alpha_{+}}$ yields
\begin{equation}
    \mel{u^\alpha_{+}}{\partial_{i}\widetilde{H}_\alpha}{u^\alpha_{-}}
    = -\Delta(k)\braket{u^\alpha_{+}}{\partial_{i}u^\alpha_{-}}.
    \label{eq:HF-app}
\end{equation}
The required derivatives of $\widetilde{H}$ are
\begin{equation}
    \partial_{k_x}\widetilde{H}_\alpha = \begin{pmatrix} 2a k_x & \alpha c \\ \alpha c & 0 \end{pmatrix},\quad
    \partial_{k_y}\widetilde{H}_\alpha = \begin{pmatrix} 2a k_y & i c \\ -i c & 0 \end{pmatrix}.
    \label{eq:dH-app}
\end{equation}
Substituting the explicit states Eq.~\eqref{eq:bloch-states-app} (and using $k_{-\alpha}^{*} = k_{\alpha}$):
\begin{align}
    (\partial_{k_x}\widetilde{H}_\alpha)\ket{u^\alpha_{-}}
    &= \frac{1}{\sqrt{N}}\begin{pmatrix} -2ack_x + ca k_{-\alpha} \\ -\alpha c^2 \end{pmatrix}
     = \frac{1}{\sqrt{N}}\begin{pmatrix} -ca k_{\alpha} \\ -\alpha c^{2} \end{pmatrix}
     = -\alpha c\ket{u^\alpha_{+}},
    \label{eq:dHum-x}
\end{align}
and similarly $(\partial_{k_y}\widetilde{H}_\alpha)\ket{u^\alpha_{-}} = +\ii c\,\ket{u^\alpha_{+}}$, so that
\begin{equation}
    \mel{u^\alpha_{+}}{\partial_{k_x}\widetilde{H}_\alpha}{u^\alpha_{-}} = -\alpha c,\qquad
    \mel{u^\alpha_{+}}{\partial_{k_y}\widetilde{H}_\alpha}{u^\alpha_{-}} = +\ii c.
    \label{eq:M-matrix}
\end{equation}
Inserting into Eq.~\eqref{eq:HF-app} gives
\begin{equation}
    \braket{u^\alpha_{+}}{\partial_{k_x}u^\alpha_{-}} = \alpha\frac{c}{\Delta(k)},\qquad
    \braket{u^\alpha_{+}}{\partial_{k_y}u^\alpha_{-}} = -i \frac{c}{\Delta(k)}.
    \label{eq:ME-app}
\end{equation}
(The step $-2ack_x + ca k_{-\alpha} = ca(k_x-\ii\alpha k_y - 2k_x) = -ca(k_x+\ii\alpha k_y) = -ca k_{\alpha}$ in Eq.~\eqref{eq:dHum-x}.)
Inserting into Eq.~\eqref{eq:QGT-app} yields the components
\begin{align}
    Q^\alpha_{xx} & = \frac{c^{2}}{\Delta(k)^{2}},\quad
    Q^\alpha_{yy} = \frac{c^{2}}{\Delta(k)^{2}},\quad
    Q^\alpha_{xy} = -i\alpha\frac{c^{2}}{\Delta(k)^{2}}.
\end{align}
Therefore the quantum metric $\metric^\alpha_{ij}=\mathrm{Re}\,Q^\alpha_{ij}$ and Berry curvature $\Berry^\alpha=-2\,\mathrm{Im}\,Q^\alpha_{xy}$ are
\begin{equation}
    \metric^\alpha_{xx} = \metric^\alpha_{yy} = \frac{c^{2}}{\Delta(k)^{2}},\quad
    \metric^\alpha_{xy}=0,\quad
    \Berry^\alpha = \alpha\frac{2c^{2}}{\Delta(k)^{2}},
    \label{eq:g-and-Omega}
\end{equation}
and thus we find the ideal quantum geometry condition
\begin{eqnarray}
    \tr\metric^\alpha = \frac{2c^{2}}{\Delta(k)^{2}} = |\Berry^\alpha|
    \label{eq:trg-Omega}
\end{eqnarray}
for either spin sector. Substituting $\Delta(k)=a k^{2}+c^{2}/a$ reproduces Eq.~\eqref{eq:geometry-results} of the main text. 

\subsection{Integrated quantum metric and quantum distance}
We consider the momentum cut-off $\Lambda$ and integration the Berry curvature over the disk $|\kk|\le\Lambda$ gives
\begin{align}
    T & = \frac{1}{2\pi}\int_{0}^{\Lambda} \frac{2c^{2}}{(a k^{2}+c^{2}/a)^{2}}\,k\,\dd k.
    \label{eq:T-integral-app}
\end{align}
Substituting $\nu = a k^{2} + c^{2}/a$ (so $\dd\nu = 2ak\,\dd k$, i.e.\ $k\,\dd k = \dd\nu/(2a)$):
\begin{align}
    T
    &= \frac{c^{2}}{2\pi a}\int_{c^{2}/a}^{a\Lambda^{2}+c^{2}/a} \nu^{-2}\,\dd\nu
    = \frac{c^{2}}{2\pi a}\left[-\frac{1}{\nu}\right]_{c^{2}/a}^{a\Lambda^{2}+c^{2}/a}
    \nonumber \\
    &= \frac{c^{2}}{2\pi a}\left(\frac{a}{c^{2}} - \frac{1}{a\Lambda^{2}+c^{2}/a}\right)
    = \frac{a\Lambda^{2}}{2\pi(a\Lambda^{2} + c^{2}/a)}.
    \label{eq:T-app}
\end{align}
In the $\Lambda\to\infty$ limit, $2\pi T\to 1$, matching the magnitude of the Chern number for either sector.
For two normalized Bloch states, we define the quantum distance by
$d^{2}(\kk,\kk') \equiv 1-|\langle u^\alpha_{-}(\kk')|u^\alpha_{-}(\kk)\rangle|^{2}$.
Taking $\kk'=0$, the sector-independent distance follows from
\begin{equation}
    |\langle u^\alpha_{-}(0)|u^\alpha_{-}(\kk)\rangle|^{2}
    = \frac{c^{2}}{a^{2}k^{2}+c^{2}},\qquad
    d^{2}(\kk,0) = 1 - \frac{c^{2}}{a^{2}k^{2}+c^{2}} = \frac{a^{2}k^{2}}{a^{2}k^{2}+c^{2}},
    \label{eq:distance-app}
\end{equation}
where we used $\ket{u^\alpha_{-}(0)} = (-1,0)^{T}$ (from Eq.~\eqref{eq:bloch-states-app} at $\kk=0$).
At the cutoff edge $d_{\max} = \sqrt{a^{2}\Lambda^{2}/(a^{2}\Lambda^{2}+c^{2})} = \sqrt{a\Lambda^{2}/(a\Lambda^{2}+c^{2}/a)} = \sqrt{2\pi T}$.

\section{Landau-Level Spectrum Diagnostics}
\label{app:LL-spectrum}

\subsection{Exact Landau level spectrum}

We provide a compact derivation of the Landau level (LL) energies for completeness.
In the symmetric gauge $\mathbf{A} = \frac{\mathcal{B}}{2}(-y,x,0)$ and with $B = e\mathcal{B}/\hbar$, the kinetic momentum operators are
\begin{equation}
    \Pi_x = -\ii\partial_x - \frac{B}{2}y,
    \Pi_y = -\ii\partial_y + \frac{B}{2}x,\qquad [\Pi_x,\Pi_y] = -\ii B.
    \label{eq:Pi-app}
\end{equation}
Defining $\Pi_{\pm} = \Pi_x \pm \ii\Pi_y$, we have $[\Pi_-,\Pi_+] = 2B$.
The cyclotron ladder operators $b = \Pi_-/\sqrt{2B}$, $b^{\dagger} = \Pi_+/\sqrt{2B}$ satisfy $[b,b^{\dagger}]=1$, and $\Pi_x^{2}+\Pi_y^{2} = 2B(b^{\dagger}b+\tfrac12)$ so that $a(\Pi_x^2+\Pi_y^2) = aB(2\hat n+1)$ with $\hat n = b^\dagger b$.
In this basis, the shifted Hamiltonians $\widetilde H_\alpha=H_\alpha-E_0\mathbb I$ take the form
\begin{equation}
    \widetilde H_{\uparrow}
    =
    \begin{pmatrix}
        aB(2\hat n+1) & c\sqrt{2B}\,b^\dagger \\
        c\sqrt{2B}\,b & c^2/a
    \end{pmatrix},
    \qquad
    \widetilde H_{\downarrow}
    =
    \begin{pmatrix}
        aB(2\hat n+1) & -c\sqrt{2B}\,b \\
        -c\sqrt{2B}\,b^\dagger & c^2/a
    \end{pmatrix}.
    \label{eq:ladder-H-app}
\end{equation}

\paragraph{Spin-up ($H_{\uparrow}$) sector.}
The Hamiltonian $\widetilde H_{\uparrow}$ acts on the two-component spinor $\ket{\Psi} = (u\ket{n,m},\,v\ket{n-1,m})^{T}$ for $n\ge 1$.
Using $\Pi_+\ket{n-1,m} = \sqrt{2Bn}\ket{n,m}$ and $\Pi_-\ket{n,m} = \sqrt{2Bn}\ket{n-1,m}$, the eigenvalue equation reduces to the $2\times2$ block
\begin{equation}
    \begin{pmatrix}
        (2n+1)aB    & c\sqrt{2Bn} \\
        c\sqrt{2Bn} & c^{2}/a
    \end{pmatrix}
    \begin{pmatrix}u\\v\end{pmatrix}
    = (\lambda-E_0)
    \begin{pmatrix}u\\v\end{pmatrix}.
    \label{eq:LL-block-app}
\end{equation}
The characteristic equation reads $\lambda'^{2} - U_n \lambda' + c^{2}B = 0$ with $\lambda' = \lambda-E_0$ and $U_n = c^{2}/a + (2n+1)aB$, giving $\lambda' = \tfrac{1}{2}[U_n \pm \sqrt{U_n^2-4c^2B}]$, which yields Eq.~\eqref{eq:ELL-energies} of the main text.

The $n=0$ zero mode for spin-up sector follows because $\Pi_-\ket{0,m}=0$ annihilates the off-diagonal term and only $u\ket{0,m}$ survives with eigenvalue $E_0+aB$, giving $\ELLzero(B)=E_0+aB$.

\paragraph{Spin-down ($H_{\downarrow}$) sector.}
For $\widetilde H_{\downarrow}$ (time-reversed sector), the natural spinor is $(\tilde u\ket{n-1,m}, \tilde v\ket{n,m})^{T}$.
After absorbing a relative minus sign into one spinor component, the eigenvalue equation reduces to the analogous block
\begin{equation}
    \begin{pmatrix}
        (2n-1)aB    & c\sqrt{2Bn} \\
        c\sqrt{2Bn} & c^{2}/a
    \end{pmatrix}
    \begin{pmatrix}\tilde u\\\tilde v\end{pmatrix}
    = (\lambda-E_0)
    \begin{pmatrix}\tilde u\\\tilde v\end{pmatrix}.
    \label{eq:LLTR-block-app}
\end{equation}
with characteristic equation $\lambda'^{2} - V_n \lambda' - c^{2}B = 0$ where $V_n = c^{2}/a + (2n-1)aB$. 
This gives $\lambda' = \tfrac{1}{2}[V_n \pm \sqrt{V_n^2+4c^2B}]$, yielding Eq.~\eqref{eq:ELLTR-energies} of the main text.
The $n=0$ zero mode for spin-down sector is $\ket{\Psi} = (0,\ket{0,m})^{T}$ with $\ELLTRzero = E_0+c^{2}/a = E_1$ (field-independent).

\subsection{LL Gap scaling asymptotics}
Here we focus on the spin-up sector and discuss the gap scaling of the anomalous LL manifold $\ELLminus(n)$ with $n=1,\ldots,\Lmax$, while the LL gap scaling for the spin-down sector $\ELLTRminus(n)$ is similar.
We distinguish the top gap, which involves the zero mode, from the gaps internal to the anomalous manifold:
\begin{equation}
    G^\uparrow_{0}=\ELLzero-\ELLminus(1),\qquad
    G^\uparrow_{n}=\ELLminus(n)-\ELLminus(n+1),\quad n=1,\ldots,\Lmax-1.
    \label{eq_SM:LL_gap_def}
\end{equation}

For $U_n^{2}\gg 4c^{2}B$ (valid when $B\ll\Bstar \equiv c^2/a^2$ or $n$ is large), one expands the square root:
$\sqrt{U_n^2-4c^2B} \approx U_n - 2c^2B/U_n$, giving 
\begin{eqnarray}
   \ELLminus(n) \approx E_0 + c^2B/U_n. 
   \label{eq_SM:ELLminus-approx}
\end{eqnarray}
From Eq.~\eqref{eq_SM:LL_gap_def}, we have:
\begin{itemize}
    \item \textbf{Internal gaps}: for adjacent anomalous levels in the small-$B$ or large-$n$ regime,
          \begin{equation}
              G^\uparrow_n
              \approx c^2B\left(\frac{1}{U_n}-\frac{1}{U_{n+1}}\right)
              = \frac{2ac^{2}B^{2}}{U_nU_{n+1}},
              \qquad
              U_n=\frac{c^2}{a}+(2n+1)aB .
              \label{eq:G-bulk-app}
          \end{equation}
          For fixed $n$ and $B\to0$, this reduces to 
          \begin{equation}
            G^\uparrow_n\approx 2a^3B^2/c^2. 
            \label{eq:G-bulk-lowB}
          \end{equation}
          
    \item \textbf{Top gap} ($n=0$): the exact gap between $\ELLzero$ and $\ELLminus(1)$ is
          \begin{equation}
              G^\uparrow_{0} = -\frac12(aB+\frac{c^{2}}{a}) + \frac12\sqrt{(\frac{c^{2}}{a})^{2}+2c^{2}B+9a^{2}B^{2}}.
              \label{eq:Gtop-exact}
          \end{equation}
          (Derivation: $G^\uparrow_{0} = (E_0+aB) - [E_0 + \frac{1}{2}(U_1-\sqrt{U_1^2-4c^2B})]$ with $U_1=c^2/a+3aB$; one simplifies $(c^2/a+3aB)^2-4c^2B = (c^2/a)^2+2c^2B+9a^2B^2$.)
          For $B\ll\Bstar$ ($a^{2}B\ll c^{2}$), $G^\uparrow_{0}\approx 2a^{3}B^{2}/c^{2}$.
          For $B\gg\Bstar$ ($a^{2}B\gg c^{2}$), $G^\uparrow_{0}\approx aB - c^{2}/(3a)$.
\end{itemize}

\begin{figure}[htbp]
    \centering
    \includegraphics[width=\linewidth]{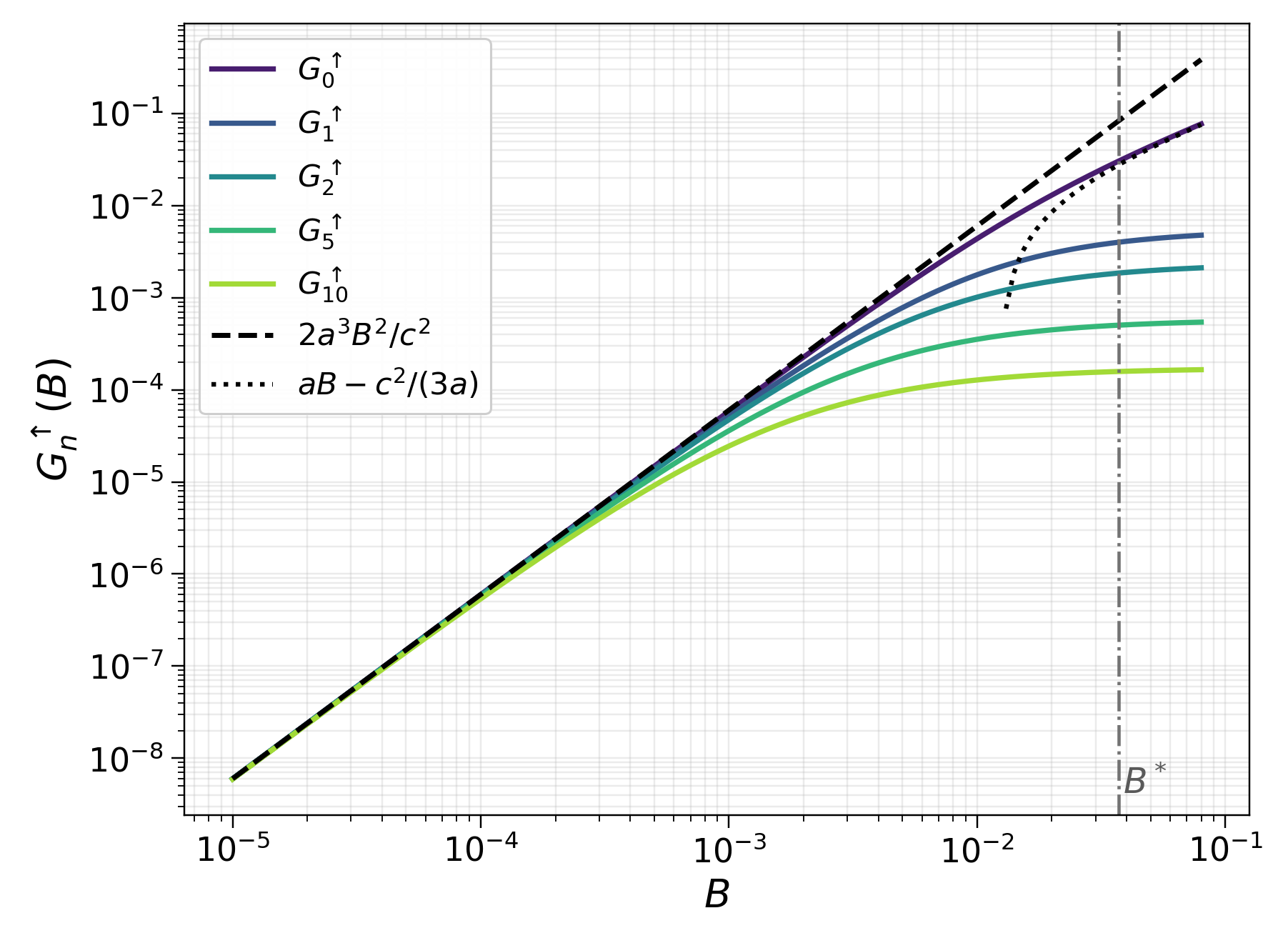}
    \caption{Exact fixed-index gap scaling from Eq.~\eqref{eq_SM:LL_gap_def} for the spin-up anomalous LL manifold. The dashed line shows the fixed-index low-field asymptote $2a^3B^2/c^2$, while the dotted line shows the large-field asymptote $aB-c^2/(3a)$ for the top gap. The vertical dash-dotted line marks $B^*=c^2/a^2$.}
    \label{fig:gap-scaling}
\end{figure}

For fixed $n$, the low-field limit is controlled by $(2n+1)aB\ll c^2/a$.
In this regime the gaps defined in Eq.~\eqref{eq_SM:LL_gap_def} are smooth functions of $B$ and scale as $B^2$ according to Eq.~\eqref{eq:G-bulk-lowB}, with an $n$-dependent crossover scale.
For a fixed carrier density away from the continuum cutoff, the relevant spacing is obtained by evaluating Eq.~\eqref{eq:G-bulk-app} near the density-selected index $n^*(\rho,B)$.
Fig.~\ref{fig:gap-scaling} plots the LL gap spacings defined in Eq.~\eqref{eq_SM:LL_gap_def}.

\section{Branch-Resolved LK Formulas}
\label{app:lk-branch}
\label{app:LL-spacing-formulas}

\subsection{Poisson-summation and contour evaluation}
\label{app:LK-derivation}

We outline the derivation of Eq.~\eqref{eq:omega-tilde} and Eq.~\eqref{eq:Xp-manuscript} of the main text from the exact LL spectrum.
For a LL branch labelled by spin sector $\alpha$ and band index $\eta=\pm$, with energies $E^\alpha_\eta(n,B)$ and degeneracy per unit area $D=B/(2\pi)$, the grand potential per unit area is
\begin{equation}
    \Omega(B,T,\mu)
    =
    -\kB T D \sum_{\alpha,\eta} \sum_{n=0}^{\Lmax}
    \ln\!\bigl[1+e^{-(E^\alpha_\eta(n,B)-\mu)/\kB T}\bigr].
    \label{eq:grand-potential-branch-app}
\end{equation}
Applying Poisson summation to this sum isolates the oscillatory part as
\begin{eqnarray}
    && \delta\Omega(B,T,\mu) = \sum_{\alpha,\eta} \delta\Omega_{\alpha,\eta}(B,T,\mu), \nonumber \\
    && \delta\Omega_{\alpha,\eta} (B,T,\mu) = -2 \kB T D\,\Re\sum_{p=1}^{\infty}
    \int_{-1/2}^{\Lmax+1/2} dn\,
    \ln \bigl[1+e^{-(E^\alpha_\eta(n,B)-\mu)/\kB T}\bigr]\,
    e^{2\pi\ii p n},
    \label{eq:omega-osc-raw}
\end{eqnarray}
with $D = B/(2\pi)$ and $\Re$ takes the real part.
Here the index $n$ is treated as a continuous variable, and the upper limit $\Lmax$ is set by the continuum cutoff.
Starting from this form, change variable to $\varepsilon=E^\alpha_\eta(n,B)$ with $dn=d\varepsilon/v_{\varepsilon}$, where 
\begin{eqnarray}
    v_{\varepsilon}=|\partial E^\alpha_\eta(n,B)/\partial n|_{n(\varepsilon)}
\end{eqnarray}
is the local LL spacing.
The temperature-dependent part of the integral is controlled by energies within a window of order $\kB T$ around the chemical potential.
When the local spacing varies slowly over this window, one may expand the inverse LL index $n_{\alpha,\eta}(\varepsilon,B)$ about the Fermi crossing $n^*_{\alpha,\eta}$ defined by 
\begin{eqnarray}
    E^\alpha_\eta(n^*_{\alpha,\eta},B)=\mu.    
\end{eqnarray}
Keeping the leading term gives
\begin{equation}
    n_{\alpha,\eta}(\varepsilon,B) \approx n^{*}_{\alpha,\eta} + \frac{\varepsilon-\mu}{v^{\alpha,\eta}_{\mu}},\qquad
    v^{\alpha,\eta}_\mu \equiv \bigl|\partial E^\alpha_\eta(n,B)/\partial n\bigr|_{n^*_{\alpha,\eta}},
\label{eq:vmu-def-app}
\end{equation}
where $v^{\alpha,\eta}_\mu$ is equivalent to the definition in Eq.~\eqref{eq:omega-tilde} of the main text.
Taking $v_{\varepsilon}\approx v^{\alpha,\eta}_{\mu}$ outside the integral gives
\begin{equation}
    \delta\Omega_{\alpha,\eta} \approx -\frac{2\kB T D}{v^{\alpha,\eta}_{\mu}}
    \Re\sum_{p=1}^{\infty} e^{2\pi\ii p n^{*}_{\alpha,\eta}}
    \int d\varepsilon
    \ln\!\bigl[1+e^{-(\varepsilon-\mu)/\kB T}\bigr]
    e^{2\pi\ii p (\varepsilon-\mu)/v^{\alpha,\eta}_{\mu}}.
    \label{eq:omega-after-approx}
\end{equation}

Due to the logarithm in Eq.~\eqref{eq:omega-after-approx}, we first need to consider the integral boundary. For $\xi=(\varepsilon-\mu)/(\kB T)\to-\infty$, $\ln(1+e^{-\xi})\sim-\xi$, so the boundary term from a direct integration by parts would not vanish.
Let $F(\xi)=\ln(1+e^{-\xi})$, $\tilde v = 2\pi p\kB T/v^{\alpha,\eta}_\mu$, and keep the finite dimensionless endpoints $\xi_\pm=(\varepsilon_\pm-\mu)/(\kB T)$ inherited from the LL cutoff.
For this finite interval, two integrations by parts give the exact identity
\begin{equation}
    \int_{\xi_-}^{\xi_+}d\xi\,F(\xi)e^{\ii\tilde v\xi}
    = {\cal B}_{\rm end}
    -\frac{1}{\tilde v^2}
    \int_{\xi_-}^{\xi_+}d\xi\,F''(\xi)e^{\ii\tilde v\xi},
    \qquad
    F''(\xi)=\frac{1}{4}\operatorname{sech}^2\!\left(\frac{\xi}{2}\right).
\end{equation}
Here ${\cal B}_{\rm end}$ contains only finite-endpoint terms evaluated at $\xi_\pm$.
These terms are cutoff/edge contributions, whereas the LK oscillation associated with the Fermi crossing is controlled by the nonanalytic change of occupation at $\varepsilon=\mu$ at zero temperature.
When the Fermi level is far from the endpoints on the scale of $\kB T$, the Fermi-surface contribution can therefore be isolated by extending the localized kernel $F''(\xi)$ to the full real line.
In the $T\to0$ limit, $F''(\xi)$ tends to $\delta(\xi)$, giving the undamped $T=0$ harmonic; at finite $T$ the broadened kernel gives the LK damping factor
\begin{align}
    R_T(\tilde v)
    &=
    \int_{-\infty}^{\infty} d\xi\,
    \left[-f'_0(\xi)\right]e^{\ii\tilde v\xi}
    =
    \int_{-\infty}^{\infty} \frac{d\xi}{4}
    \operatorname{sech}^2\!\!\left(\frac{\xi}{2}\right)e^{\ii\tilde v\xi},
\end{align}
where $f_0(\xi) = (1+e^\xi)^{-1}$. In this expression for $R_T(\tilde v)$, the imaginary part vanishes because the thermal kernel is even in $\xi$, whereas $\sin(\tilde v\xi)$ is odd.
Closing the contour in the upper half-plane picks up the Matsubara poles of $f'_0(\xi)$ at $\xi_k = \ii\pi(2k+1)$ ($k=0,1,2,\ldots$):
\begin{equation}
    R_T(\tilde v)
    = \frac{X^{\alpha,\eta}_p}{\sinh(X^{\alpha,\eta}_p)}.
\end{equation}
where 
\begin{eqnarray}
    X^{\alpha,\eta}_p \equiv \pi\tilde v = 2\pi^2 p \kB T/v^{\alpha,\eta}_\mu.
\end{eqnarray}
Collecting all prefactors -- the Jacobian $1/v^{\alpha,\eta}_\mu$ from the change of variable, two integration-by-parts factors of $v^{\alpha,\eta}_\mu/(2\pi p)$, and the explicit $D$ factor -- yields
\begin{equation}
    \delta\Omega_{\alpha,\eta}(B,T,\mu) = \frac{D}{2\pi^{2}}\sum_{p=1}^{\infty}\frac{v^{\alpha,\eta}_{\mu}}{p^{2}}
    \,\frac{X^{\alpha,\eta}_p}{\sinh X^{\alpha,\eta}_p}\,\cos\!\bigl[2\pi p\, n^{*}_{\alpha,\eta}(\mu, B)\bigr],
    \label{eq:branch-lk-omega-app}
\end{equation}
in agreement with Eq.~\eqref{eq:omega-tilde} of the main text.
(For complete bookkeeping, see e.g.\ Shoenberg, \emph{Magnetic Oscillations in Metals}, \S2.3, eqs.\ 2.151--2.157.)

\subsection{Local LL spacing for each branch}

\paragraph{$\ELLminus$ branch ($\mu>E_0$, wedge condition $0<\dmu<c\sqrt{B}$).}
Setting $\ELLminus(n^*,B) = \mu$ with $\dmu = \mu-E_0>0$ means
\begin{equation}
    2\dmu = U_{n^{*}} - \sqrt{U_{n^{*}}^{2} - 4c^{2}B}.
    \label{eq:cross-LL-app}
\end{equation}
Rearranging to isolate the square root and squaring:
$(U_{n^*}-2\dmu)^2 = U_{n^*}^2 - 4c^2B$, which gives $4\dmu^2 - 4\dmu U_{n^*} = -4c^2B$, hence
\begin{equation}
    U_{n^{*}} \equiv \frac{c^{2}}{a}+(2n^{*}+1)aB
    = \dmu + \frac{c^{2}B}{\dmu}.
    \label{eq:U-star-app}
\end{equation}
(The squaring step requires $U_{n^*}>2\dmu$, i.e.\ $c^2B>\dmu^2$, which is precisely the wedge condition.)
Solving Eq.~\eqref{eq:U-star-app} for $n^*$:
\begin{equation}
    n^{*} + \frac12
    = \frac{U_{n^*}-c^2/a}{2aB}
    = -\frac{E_0+c^2/a-\mu}{2aB} + \frac{c^{2}}{2a(\mu-E_0)}.
    \label{eq:nstar-LL-app}
\end{equation}
Here the combination $E_0+c^2/a$ enters only through the ideal-flat-band offset in the spin-up block.
The local spacing follows from $\partial E_{-}/\partial n\big|_{n^*}$. Differentiating $\ELLminus = E_0 + \tfrac12(U_n - \sqrt{U_n^2-4c^2B})$ with respect to $n$ (using $\partial U_n/\partial n = 2aB$) and evaluating at $n^*$ where $\sqrt{U_{n^*}^2-4c^2B} = U_{n^*}-2\dmu$:
\begin{equation}
    \frac{\partial \ELLminus}{\partial n}\bigg|_{n^{*}}
    = aB\!\left(1 - \frac{U_{n^*}}{U_{n^*}-2\dmu}\right)
    = \frac{-2aB\,\dmu}{U_{n^*}-2\dmu}
    = \frac{-2aB\,\dmu^2}{c^2B-\dmu^2},
\end{equation}
where we used $U_{n^*}-2\dmu = (c^2B-\dmu^2)/\dmu$ from Eq.~\eqref{eq:U-star-app}.
Since $\dmu < c\sqrt{B}$, the denominator $c^2B-\dmu^2>0$ and the slope is negative (levels decrease in $n$).
Taking the absolute value of this local slope gives the positive LL spacing:
\begin{equation}
    v_{\mu}^{\uparrow,-} \equiv \Bigl|\frac{\partial \ELLminus}{\partial n}\Bigr|_{n^{*}}
    = \frac{2aB(\mu-E_0)^{2}}{c^{2}B - (\mu-E_0)^{2}}.
    \label{eq:dEdn-LLminus-app}
\end{equation}

\paragraph{$\ELLTRminus$ branch ($\mu<E_0$, valid for $-c\sqrt{B}<\dmu<0$).}
Setting $\ELLTRminus(n^*,B) = \mu$ with $\dmu = \mu-E_0<0$:
\begin{equation}
    2\dmu = V_{n^{*}} - \sqrt{V_{n^{*}}^{2} + 4c^{2}B}.
    \label{eq:cross-LLTR-app}
\end{equation}
Isolating the square root and squaring: $(V_{n^*}-2\dmu)^2 = V_{n^*}^2+4c^2B$, giving
$4\dmu^2 - 4\dmu V_{n^*} = 4c^2B$, hence (with $\dmu<0$)
\begin{equation}
    V_{n^{*}} \equiv \frac{c^{2}}{a}+(2n^{*}-1)aB
    = \dmu - \frac{c^{2}B}{\dmu}.
    \label{eq:V-star-app}
\end{equation}
Solving for $n^*$:
\begin{equation}
    n^{*} - \frac12
    = \frac{V_{n^*}-c^2/a}{2aB}
    = \frac{c^{2}}{2a(E_0-\mu)} - \frac{E_0+c^2/a-\mu}{2aB}.
    \label{eq:nstar-LLTR-app}
\end{equation}
The local spacing can be obtained directly from the slope of the spin-down anomalous branch.
Differentiating $\ELLTRminus = E_0+\tfrac12(V_n-\sqrt{V_n^2+4c^2B})$ with respect to $n$ and using $\partial V_n/\partial n=2aB$ gives
\begin{equation}
    \frac{\partial \ELLTRminus}{\partial n}\bigg|_{n^*}
    =
    aB\!\left(1-\frac{V_{n^*}}{\sqrt{V_{n^*}^2+4c^2B}}\right).
\end{equation}
At the Fermi crossing, Eq.~\eqref{eq:cross-LLTR-app} gives
$\sqrt{V_{n^*}^2+4c^2B}=V_{n^*}-2\dmu$.
Using Eq.~\eqref{eq:V-star-app},
\begin{equation}
    V_{n^*}-2\dmu
    =
    -\dmu-\frac{c^2B}{\dmu}
    =
    -\frac{\dmu^2+c^2B}{\dmu},
\end{equation}
which is positive because $\dmu<0$.
Therefore
\begin{equation}
    \frac{\partial \ELLTRminus}{\partial n}\bigg|_{n^*}
    =
    aB\left[1+
    \frac{\dmu^2-c^2B}{\dmu^2+c^2B}\right]
    =
    \frac{2aB\dmu^2}{\dmu^2+c^2B}.
\end{equation}
The slope is positive for this branch, so the positive local spacing is
\begin{equation}
    v_{\mu}^{\downarrow,-} \equiv \Bigl|\frac{\partial \ELLTRminus}{\partial n}\Bigr|_{n^{*}}
    = \frac{2aB(\mu-E_0)^{2}}{(\mu-E_0)^{2} + c^{2}B}.
    \label{eq:dEdn-LLTR-app}
\end{equation}

\paragraph{Semiclassical (bulk) limit and connection to quantum geometry.}
In the bulk of the anomalous manifold ($|\dmu|\ll c\sqrt{B}$, large $n^*$), Eq.~\eqref{eq:dEdn-LLminus-app} first reduces to
\begin{equation}
    v_{\mu}^{\uparrow,-}
    =\frac{2aB\dmu^2}{c^2B-\dmu^2}
    \approx \frac{2a\dmu^2}{c^2}.
\end{equation}
To relate $\dmu$ to the zero-field band geometry, use the large-$n^*$ semiclassical identification $k_{n^*}^2\approx 2n^*B$.
In the same regime the anomalous LL energy follows from expanding the lower branch as
\begin{equation}
    \dmu = \ELLminus(n^*,B)-E_0
    \approx \frac{c^2B}{U_{n^*}}
    \approx \frac{c^2B}{\Delta(k_{n^*})},
    \qquad
    \Delta(k)=ak^2+\frac{c^2}{a},
\end{equation}
where we have used Eq.\eqref{eq_SM:ELLminus-approx}. 
Substituting this estimate for $\dmu$ into the bulk expression for $v_\mu$ gives
\begin{equation}
    v_{\mu}^{\uparrow,-} \approx \frac{2a\dmu^2}{c^2}
    \approx \frac{2ac^2B^2}{\Delta(k_{n^*})^2}
    = aB^2\,\tr\metric(k_{n^*}),
    \label{eq:v-geom-app}
\end{equation}
where the last equality uses $\tr\metric = 2c^2/\Delta^2$ from Eq.~\eqref{eq:g-and-Omega}.
The same relation holds for the $\ELLTRminus$ branch. Thus the local LL spacing directly encodes the quantum metric of the flat band, establishing $v^{\alpha,-}_\mu \sim aB^2 \tr\metric$ as the fundamental connection between anomalous-LL thermal damping and quantum geometry.

\paragraph{Upper branches ($\ELLplus$, $\ELLTRplus$).}
For the upper branches, the same differentiation of the exact LL energies gives
\begin{equation}
    v_{\mu}^{\uparrow,+}
    =
    \frac{2aB(\mu-E_0)^{2}}{(\mu-E_0)^{2}-c^{2}B},
    \qquad
    v_{\mu}^{\downarrow,+}
    =
    \frac{2aB(\mu-E_0)^{2}}{(\mu-E_0)^{2}+c^{2}B}.
    \label{eq:dEdn-upper-app}
\end{equation}
In the dispersive regime $|\mu-E_0|\gg c\sqrt{B}$, both upper-branch spacings reduce to $v^{\alpha,+}_{\mu}\approx 2aB$, recovering the standard parabolic-band LK factor $X^{\alpha,+}_p = \pi^{2} p \kB T/(aB)$ with effective mass $m^*_{\mathrm{eff},\alpha +} = 1/(2a)$.

\subsection{Effective mass formulas}

Using $\meff = B/v_{\mu}$ and the above expressions for $v_{\mu}$ gives:
\begin{align}
    m^{*}_{\mathrm{eff},\uparrow-}   & = \frac{c^{2}B - (\mu-E_0)^{2}}{2a(\mu-E_0)^{2}}, \label{eq:meff-LLminus-app}  \\
    m^{*}_{\mathrm{eff},\downarrow-} & = \frac{1}{2a} + \frac{c^{2}B}{2a(\mu-E_0)^{2}}. \label{eq:meff-LLTRminus-app}
\end{align}
In the anomalous bulk regime, $\dmu^2\ll c^2B$, the two lower-branch masses have the same leading behavior,
\begin{equation}
    m^*_{\mathrm{eff},\uparrow-}
    \simeq
    m^*_{\mathrm{eff},\downarrow-}
    \simeq
    \frac{c^2B}{2a(\mu-E_0)^2},
    \label{eq:meff-anomalous-bulk-app}
\end{equation}
up to the subleading additive constants $\mp 1/(2a)$.
This strong field dependence at fixed $\mu$ contrasts with the constant parabolic value $m^*_{\mathrm{eff},\alpha +}=1/(2a)$ of the dispersive upper branches.
At fixed carrier density, however, the field dependence must be evaluated along the self-consistent trajectory $\mu_\rho(B)$.
For a single active lower-branch sequence, the LL degeneracy is $D=B/(2\pi)$, so fixed active density $\rho_{\mathcal S}$ (defined in Sec.~D as the density counted by the active sector) selects
\begin{equation}
    n^*(B)\simeq \frac{|\rho_{\mathcal S}|}{D}
    =
    \frac{2\pi|\rho_{\mathcal S}|}{B}.
    \label{eq:nstar-fixed-density-app}
\end{equation}
The semiclassical momentum tied to this crossing is therefore
\begin{equation}
    k_{n^*}^2\simeq 2n^*B\simeq 4\pi|\rho_{\mathcal S}|,
    \label{eq:kn-fixed-density-app}
\end{equation}
which is approximately independent of $B$ at fixed density.
Using the weak-field anomalous-branch dispersion,
\begin{equation}
    \mu_\rho(B)-E_0
    \simeq
    s_\alpha\frac{c^2B}{c^2/a+a k_{n^*}^2}
    \simeq
    s_\alpha\frac{c^2}{c^2/a+4\pi a|\rho_{\mathcal S}|}\,B,
    \qquad
    s_\uparrow=+1,\quad s_\downarrow=-1,
    \label{eq:mu-fixed-density-linear-app}
\end{equation}
so the chemical-potential offset from the flat-band energy is linear in $B$ along the fixed-density trajectory.
Substituting Eq.~\eqref{eq:mu-fixed-density-linear-app} into Eq.~\eqref{eq:meff-anomalous-bulk-app} gives
\begin{equation}
    m^*_{\mathrm{eff},\alpha-}\bigl(B,\mu_\rho(B)\bigr)
    \simeq
    \frac{\left(c^2/a+4\pi a|\rho_{\mathcal S}|\right)^2}{2ac^2}\,\frac{1}{B}
    =
    \frac{1}{aB\,\tr\metric_\alpha(k_{n^*})},
    \label{eq:meff-fixed-density-inverseB-app}
\end{equation}
which is the roughly $1/B$ trend shown by the dashed theory curves in Fig.~\ref{fig:main4} of the main text.

\section{Numerical Results for Fixed-Density Thermodynamics}
\label{app:thermodynamics}

\subsection{Density convention and Fermi level}

We first fix the density convention used in all numerical plots.
The signed carrier density \(\rho\) is measured relative to the midpoint of the two zero modes,
\begin{equation}
    E_{\mathrm{ref}}(B)=\frac{\ELLzero(B)+\ELLTRzero(B)}{2}.
    \label{eq:Eref-app}
\end{equation}
Thus \(\rho>0\) denotes electron doping above \(E_{\mathrm{ref}}\), while \(\rho<0\) denotes hole doping below \(E_{\mathrm{ref}}\).
The corresponding discrete LL spectral density per unit area is
\begin{equation}
    \nu_{\mathrm{LL}}(E,B) =
    \frac{B}{2\pi}
    \left[
    \sum_{\alpha=\uparrow,\downarrow}
    \delta\!\left(E-E^\alpha_0(B)\right)
    +
    \sum_{\alpha=\uparrow,\downarrow}
    \sum_{\eta=\pm}
    \sum_{n=1}^{\Lmax}
    \delta\!\left(E-E^\alpha_\eta(n,B)\right)
    \right],
    \label{eq:LL-spectral-density-app}
\end{equation}
where \(E^\alpha_0(B)\) denotes the zero mode in spin sector \(\alpha\), and \(E^\alpha_\eta(n,B)\) denotes the branch-resolved LL energy introduced in Appendix~\ref{app:LL-spectrum}.
At finite temperature, the signed density associated with a chemical potential \(\mu\) is obtained by integrating the LL spectral density with the Fermi occupation,
\begin{equation}
    \rho(\mu,B,T)
    =
    \int dE\,
    \nu_{\mathrm{LL}}(E,B)
    \left[
    f_T(E-\mu)-\Theta\!\left(E_{\mathrm{ref}}(B)-E\right)
    \right],
    \qquad
    f_T(x)=\frac{1}{e^{x/\kB T}+1}.
    \label{eq:finiteT-density-app}
\end{equation}
The second term subtracts the reference filling with all LLs below \(E_{\mathrm{ref}}\) occupied and all LLs above \(E_{\mathrm{ref}}\) empty.
For a prescribed carrier density, the chemical potential \(\mu_\rho(B,T)\) is determined by solving \(\rho=\rho(\mu_\rho,B,T)\).
In the \(T\to0\) limit, Eq.~\eqref{eq:finiteT-density-app} reduces to the usual signed LL-counting convention relative to \(E_{\mathrm{ref}}\).

Since we need to choose a momentum cutoff $\Lambda$ for the topological flat bands, which also introduces a cutoff in the anomalous LL spectrum, it is more stable to work with the active carrier sector rather than with the full signed integral each time.
This gives three cases:
\begin{align}
    \rho>0:\quad
    &\rho_{\mathcal S}=\rho,
    &&\mathcal{S}=\{\text{electron LLs above }E_{\mathrm{ref}}\},
    \notag\\
    -\rho_{\mathrm{cap}}/2<\rho<0:\quad
    &\rho_{\mathcal S}=|\rho|,
    &&\mathcal{S}=\{\text{hole levels counted downward from }E_{\mathrm{ref}}\},
    \notag\\
    \rho<-\rho_{\mathrm{cap}}/2:\quad
    &\rho_{\mathcal S}=\rho_{\mathrm{low},e}
    =\rho_{\mathrm{cap}}-|\rho|,
    &&\mathcal{S}=\{\text{remaining electrons in the lower manifold}\}.
    \label{eq:density-regimes-app}
\end{align}
The shallow-hole description is used when the missing carriers are still between \(E_0\) and \(E_{\mathrm{ref}}\), so the chemical potential is within the upper manifold of anomalous LLs. The maximal carrier density of the two anomalous sectors together is 
\begin{equation}
    \rho_{\mathrm{cap}}=\frac{\Lambda^2}{2\pi}.
    \label{eq:rho-cap-app}
\end{equation}
Thus the shallow/deep boundary in Eq.~\eqref{eq:density-regimes-app} occurs at \(\rho_{\mathrm{cap}}/2=\Lambda^2/(4\pi)\).
For \(\Lambda=1.0\,\mathrm{nm}^{-1}\), one flat band has capacity \(0.0796\,\mathrm{nm}^{-2}\), while the combined two-sector capacity is \(\rho_{\mathrm{cap}}=0.159\,\mathrm{nm}^{-2}\); therefore \(\rho=-0.140\) and \(-0.150\,\mathrm{nm}^{-2}\) are deep-hole examples used below for numerical calculations in this combined convention.
This rewriting is only a carrier-counting convention: the oscillations are then controlled by the small number \(\rho_{\mathrm{low},e}\) of lower-manifold electrons left unremoved.
Consequently, the black reference lines in the FFT plots use \(F=2\pi|\rho|\) for the electron and shallow-hole sectors, whereas the deep-hole reference uses \(F=2\pi\rho_{\mathrm{low},e}\).
As discussed below, the \(\rho=+0.020\,\mathrm{nm}^{-2}\) magnetization
spectrum also contains a dominant low-frequency peak near \(F=\pi\rho\).
This feature correlates with the near-Fermi LL splitting structure and does
not alter the fixed-density counting convention.

For each case in Eq.~\eqref{eq:density-regimes-app}, we introduce an active-sector energy variable \(\varepsilon^{\mathcal S}_\ell(B)\).
Let \(E_\ell(B)\) denote the LL energies \(E^\alpha_0(B)\) and \(E^\alpha_\eta(n,B)\) in the original electron spectrum.
Then
\begin{equation}
    \varepsilon^{\mathcal S}_\ell(B)=
    \begin{cases}
    E_\ell(B), & \mathcal S=\{\text{electron LLs above }E_{\mathrm{ref}}\},\\
    E_{\mathrm{ref}}(B)-E_\ell(B), & \mathcal S=\{\text{hole levels counted downward from }E_{\mathrm{ref}}\},\\
    E_\ell(B), & \mathcal S=\{\text{remaining electrons in the lower manifold}\}.
    \end{cases}
    \label{eq:sector-energy-app}
\end{equation}
The sector density \(\rho_{\mathcal S}\) is therefore the active carrier density for the chosen sector.
Only after this sector energy variable is fixed do we define the sector chemical potential \(\mu_{\mathcal S}\).
It is the quantity conjugate to \(\rho_{\mathcal S}\), measured in the same energy variable as \(\varepsilon^{\mathcal S}_\ell\), and is defined implicitly by
\begin{equation}
    \rho_{\mathcal S}
    =
    D\sum_{\ell\in\mathcal S}
    f_T\!\left(\varepsilon^{\mathcal S}_\ell(B)-\mu_{\mathcal S}\right),
    \qquad
    D=\frac{B}{2\pi},
    \label{eq:sector-density-app}
\end{equation}
which is the sector version of Eq.~\eqref{eq:finiteT-density-app}.
For the electron and deep-hole sectors, \(\mu_{\mathcal S}\) is the physical electron chemical potential restricted to the retained sector.
For the shallow-hole sector, \(\mu_{\mathcal S}\) is instead the hole chemical potential in the transformed variable \(E_{\mathrm{ref}}-E_\ell\), not the original electron chemical potential \(\mu\).

\subsection{Fixed-density magnetization}

For each active carrier sector $\mathcal{S}$, the sector grand potential is
\begin{equation}
    \Omega_{\mathcal{S}}(B,T,\mu_{\mathcal{S}})
    = -\kB T\,D\sum_{\ell\in\mathcal{S}}
    \ln\!\left[1+\exp\!(-\frac{\varepsilon^{\mathcal{S}}_\ell(B)-\mu_{\mathcal{S}}}{\kB T})\right],
    \label{eq:omega-sector}
\end{equation}
Here $\mathcal{S}$, \(\varepsilon^{\mathcal{S}}_\ell(B)\), \(\rho_{\mathcal S}\), \(\mu_{\mathcal{S}}(B,T,\rho_{\mathcal{S}})\), and \(D\) are the sector quantities defined above.
The magnetization (per unit area) at fixed sector density is obtained from the Helmholtz free energy $f_{\mathcal{S}} = \Omega_{\mathcal{S}} + \mu_{\mathcal{S}}\rho_{\mathcal{S}}$:
\begin{equation}
    M_{\mathcal{S}}(B,T,\rho_{\mathcal{S}}) = -\frac{\partial f_{\mathcal{S}}}{\partial B}\bigg|_{T,\rho_{\mathcal{S}}}.
    \label{eq:M-sector}
\end{equation}
To see how the fixed-density derivative is evaluated, write the chemical potential on the fixed-density trajectory as \(\mu_{\mathcal S}(B)\), suppressing \(T\) and \(\rho_{\mathcal S}\) in the notation.
Then
\begin{align}
    \frac{d f_{\mathcal S}}{dB}\bigg|_{T,\rho_{\mathcal S}}
    &=
    \left(\frac{\partial \Omega_{\mathcal S}}{\partial B}\right)_{T,\mu_{\mathcal S}}
    +
    \left(\frac{\partial \Omega_{\mathcal S}}{\partial \mu_{\mathcal S}}\right)_{T,B}
    \frac{d\mu_{\mathcal S}}{dB}
    +
    \rho_{\mathcal S}\frac{d\mu_{\mathcal S}}{dB}
    \notag\\
    &=
    \left(\frac{\partial \Omega_{\mathcal S}}{\partial B}\right)_{T,\mu_{\mathcal S}},
    \label{eq:fixed-density-cancel-app}
\end{align}
where the last equality uses the thermodynamic identity \(\rho_{\mathcal S}=-(\partial\Omega_{\mathcal S}/\partial\mu_{\mathcal S})_{T,B}\).
Thus \(M_{\mathcal S}\) can equivalently be computed as \(-(\partial\Omega_{\mathcal S}/\partial B)_{T,\mu_{\mathcal S}}\) evaluated at \(\mu_{\mathcal S}=\mu_{\mathcal S}(B,T,\rho_{\mathcal S})\).
In the numerical evaluation, the \(B\)-derivative in Eq.~\eqref{eq:fixed-density-cancel-app} can be obtained analytically from the explicit LL energies \(E^\alpha_0(B)\) and \(E^\alpha_\eta(n,B)\) given in Appendix~\ref{app:LL-spectrum}.

\begin{figure}[htbp]
    \centering
    \includegraphics[width=\linewidth]{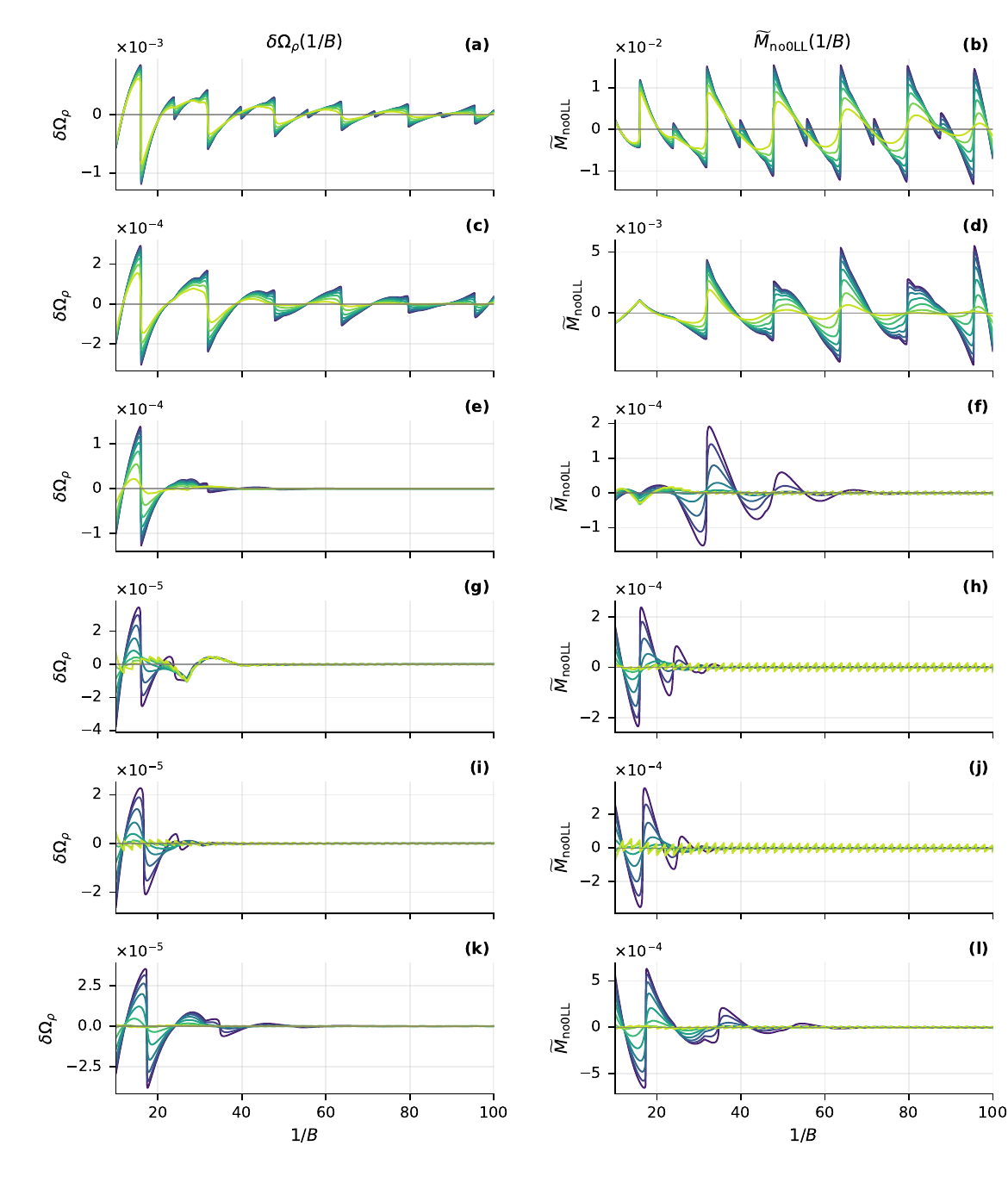}
    \caption{Wide-range thermodynamic oscillations at $\Lambda=1.0\,\mathrm{nm}^{-1}$ for $1/B\in[10,100]$. In each panel, the curves correspond to $k_B T=(2.00,\,3.07,\,4.71,\,7.22,\,11.1,\,17.0,\,26.1,\,40.0)\times10^{-4}\,\mathrm{eV}$, ordered from purple at the lowest temperature to yellow at the highest temperature. The left column shows the background-subtracted sector grand potential $\delta\Omega_{\rho}(1/B)$ computed from the exact discrete LL sum in Eq.~\eqref{eq:omega-sector}; the right column shows the background-subtracted, zero-mode-excluded oscillatory magnetization $\widetilde M_{\mathrm{no0LL}}(1/B)$ computed along the same fixed-density trajectory. The density assignments are $(a,b)$ $\rho=+0.020$, $(c,d)$ $\rho=+0.010$, $(e,f)$ $\rho=-0.010$, $(g,h)$ $\rho=-0.020$, $(i,j)$ $\rho=-0.140$, and $(k,l)$ $\rho=-0.150$, all in units of $\mathrm{nm}^{-2}$.}
    \label{fig:thermo-wide-osc}
\end{figure}

The oscillatory part $\widetilde{M}$ is obtained numerically by subtracting a smooth background in $1/B$.
For the anomalous densities, we adopt the $M_{\text{no0LL}}$ convention: the zero-mode contributions are excluded from the oscillatory signal to avoid contamination by the field-independent $\ELLTRzero$ and the trivial $B$-linear $\ELLzero$ shift, which has no contribution to oscillation behaviors.
We also evaluate the exact discrete sector grand potential in Eq.~\eqref{eq:omega-sector} along the fixed-density trajectory and subtract the same type of smooth background, denoting the resulting oscillatory component by $\delta\Omega_{\rho}$.
For deep-hole densities, the active density is the remaining lower-manifold electron density defined in Eq.~\eqref{eq:density-regimes-app}.
This construction is shown directly in Fig.~\ref{fig:thermo-wide-osc}, which plots the resulting thermodynamic oscillations over a wide \(1/B\) range for the six densities used in the appendix.
The left column shows the background-subtracted grand potential \(\delta\Omega_\rho\) obtained from the exact LL sum, while the right column shows the background-subtracted, zero-mode-excluded magnetization \(\widetilde M_{\mathrm{no0LL}}\).
Thus Fig.~\ref{fig:thermo-wide-osc} serves as the starting point for the subsequent FFT, window, and thermal-damping analyses, and it also makes the temperature dependence visible before any Fourier or local-window processing is applied. 
At the lowest temperatures, the electron densities in the dispersive regime, Figs.~\ref{fig:thermo-wide-osc}(a)--\ref{fig:thermo-wide-osc}(d), show many well-resolved oscillation periods across the plotted \(1/B\) range, whereas the hole densities in the anomalous regime, Figs.~\ref{fig:thermo-wide-osc}(e)--\ref{fig:thermo-wide-osc}(l), lose amplitude rapidly toward smaller \(B\) (larger \(1/B\)), leaving only a few visible periods before the oscillatory signal becomes strongly suppressed.

\subsection{FFT frequency diagnostics}

\begin{figure}[htbp]
    \centering
    \includegraphics[width=\linewidth]{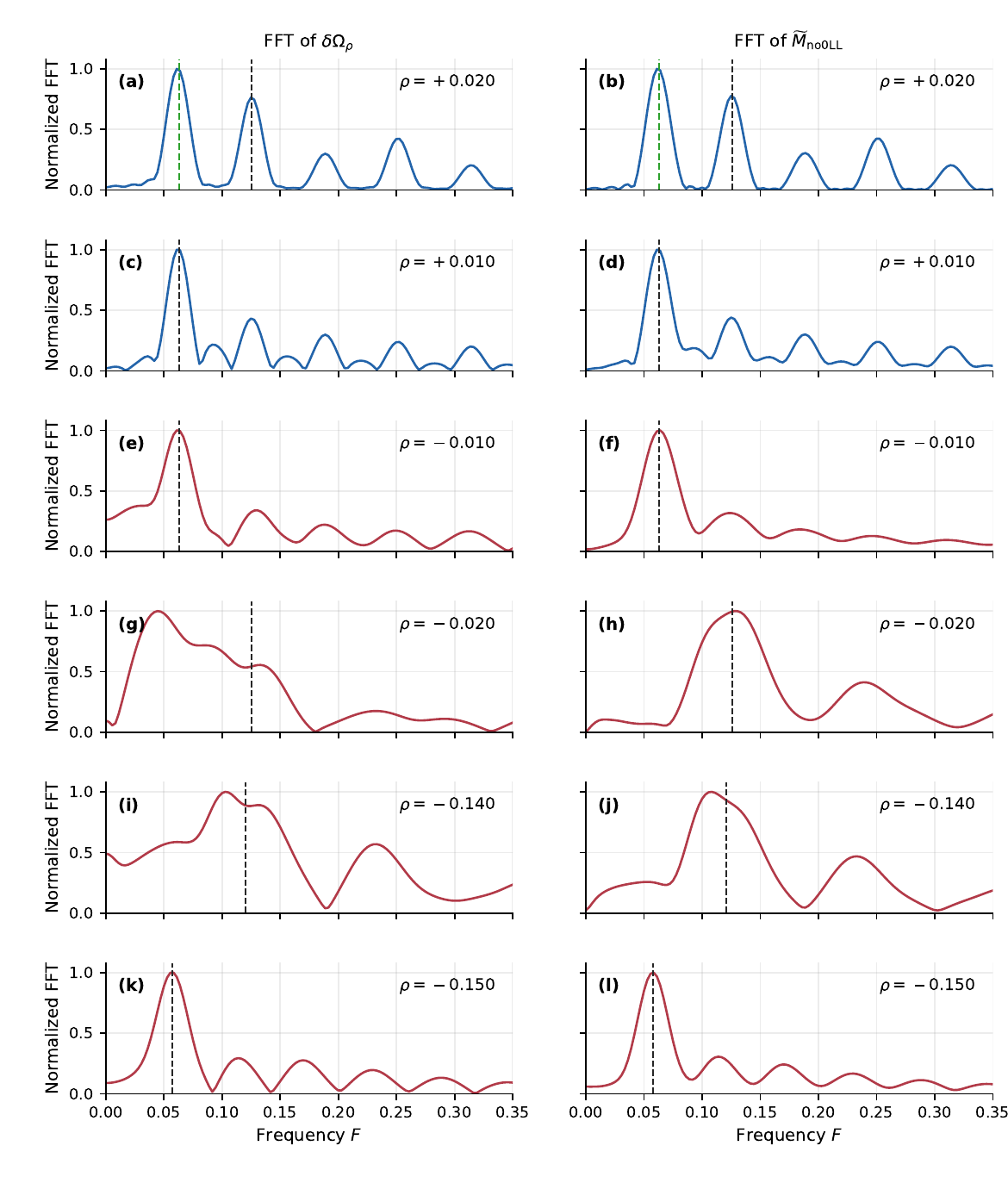}
    \caption{FFT spectra of the oscillatory grand potential $\delta\Omega_\rho(1/B)$ (left column) and zero-mode-excluded oscillatory magnetization $\widetilde M_{\mathrm{no0LL}}(1/B)$ (right column) at $\Lambda=1.0\,\mathrm{nm}^{-1}$ and $k_B T=2.0\times10^{-4}\,\mathrm{eV}$. The density assignments are $(a,b)$ $\rho=+0.020$, $(c,d)$ $\rho=+0.010$, $(e,f)$ $\rho=-0.010$, $(g,h)$ $\rho=-0.020$, $(i,j)$ $\rho=-0.140$, and $(k,l)$ $\rho=-0.150$, all in units of $\mathrm{nm}^{-2}$. Black dashed vertical lines mark the LL-counting reference $F=2\pi\rho_{\mathcal S}$, with $\rho_{\mathcal S}=|\rho|$ for the electron and shallow-hole sectors and $\rho_{\mathcal S}=\rho_{\mathrm{low},e}$ for the deep-hole sectors. The green dashed lines in panels (a) and (b) mark \(F=\pi\rho\) for \(\rho=+0.020\,\mathrm{nm}^{-2}\). }
    \label{fig:magnetization-fft}
\end{figure}

Figure~\ref{fig:magnetization-fft} is the frequency-space diagnostic of the wide-range oscillations in Fig.~\ref{fig:thermo-wide-osc}.
For each density, we Fourier transform the lowest-temperature traces, \(k_B T=2.0\times10^{-4}\,\mathrm{eV}\), using the same \(1/B\) interval as in Fig.~\ref{fig:thermo-wide-osc}.
The left column analyzes the background-subtracted grand potential \(\delta\Omega_\rho\), while the right column analyzes the background-subtracted, zero-mode-excluded magnetization \(\widetilde M_{\mathrm{no0LL}}\).
The vertical dashed lines in Fig.~\ref{fig:magnetization-fft} are LL-counting reference frequencies used to interpret where the numerical FFT weight should appear, as discussed below. 

\begin{figure}[htbp]
    \centering
    \includegraphics[width=0.86\linewidth]{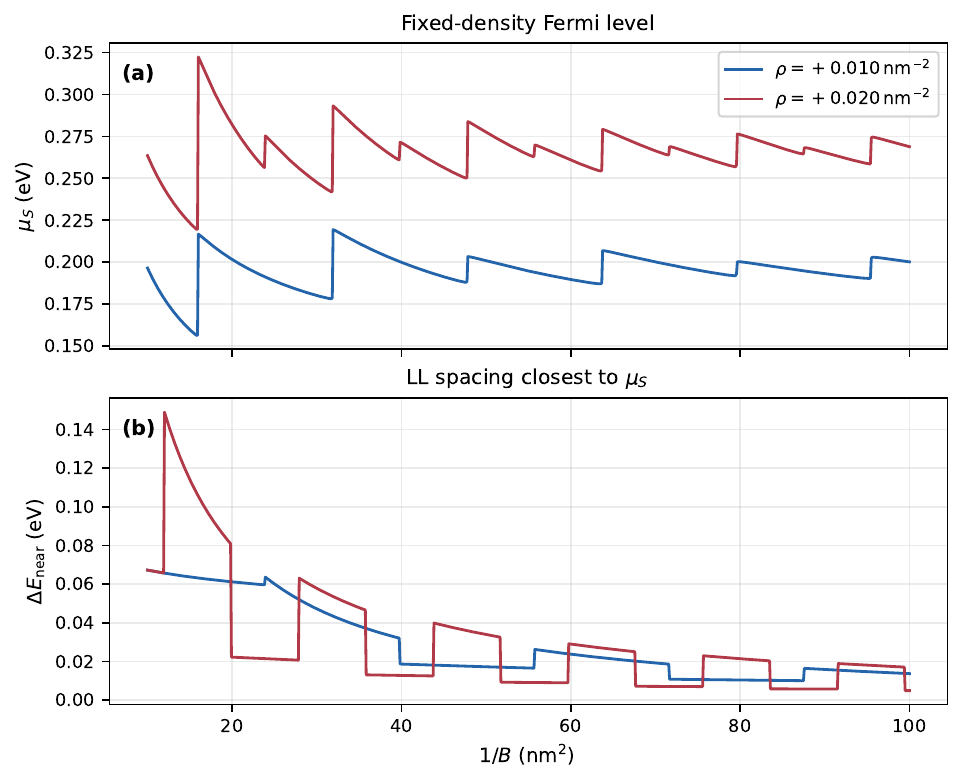}
    \caption{Comparison of the two electron-density cases used in Figs.~\ref{fig:thermo-wide-osc} and \ref{fig:magnetization-fft}. Panel (a) shows the fixed-density chemical potential \(\mu_{\mathcal S}(1/B)\) for \(\rho=+0.010\) and \(+0.020\,\mathrm{nm}^{-2}\) at \(k_B T=2.0\times10^{-4}\,\mathrm{eV}\). Panel (b) shows the nearest-chemical-potential LL spacing \(\Delta E_{\mathrm{near}}=E_{\mathrm{above}}-E_{\mathrm{below}}\), where \(E_{\mathrm{below}}<\mu_{\mathcal S}<E_{\mathrm{above}}\) are the two LLs closest to the fixed-density chemical potential. }
    \label{fig:electron-mu-splitting-comparison}
\end{figure}

For one set of LL ladder, the degeneracy per unit area is \(D=B/(2\pi)\).
If \(n_f\) levels are filled, the active carrier density is \(\rho_{\mathcal S}=D n_f\), so \(n_f=(2\pi\rho_{\mathcal S})/B\).
One oscillation occurs when \(n_f\) changes by one as a function of \(1/B\), giving the fundamental frequency
\begin{equation}
    F_{\rho_{\mathcal S}} = 2\pi\rho_{\mathcal S}.
    \label{eq:onsager-freq-app}
\end{equation}
Here \(\rho_{\mathcal S}\) is the density counted by the active LL sequence.
For the electron and shallow-hole rows, \(\rho_{\mathcal S}=|\rho|\).
For the deep-hole rows, the useful active variable is the remaining lower-manifold electron density \(\rho_{\mathcal S}=\rho_{\mathrm{low},e}\) defined in Eq.~\eqref{eq:density-regimes-app}, because the oscillations count the residual unfilled part of the lower anomalous manifold rather than the full hole density.
The black dashed lines in Fig.~\ref{fig:magnetization-fft} mark \(F=2\pi\rho_{\mathcal S}\) for both electron and hole sectors.
With this convention, the FFT spectra test whether the exact fixed-density LL calculation oscillates with the period expected from the active carrier density.
The electron-side spectra are shown first in Figs.~\ref{fig:magnetization-fft}(a)--\ref{fig:magnetization-fft}(d).
For \(\rho=+0.010\,\mathrm{nm}^{-2}\), panels (c) and (d) have their dominant peak consistent with the black line from Eq.~\eqref{eq:onsager-freq-app}, so the oscillations are well described by a spin-resolved LL ladder with \(\rho_{\mathcal S}=\rho\).
For \(\rho=+0.020\,\mathrm{nm}^{-2}\), panels (a) and (b) instead show a stronger low-frequency peak described by the green dashed line \(F=\pi\rho\), rather than by the black dashed line at \(F=2\pi\rho\).
This additional peak is consistent with an effectively spin-degenerate
density-frequency relation. If two spin ladders are sufficiently close over
the relevant field interval, the fixed total carrier density is shared between
them.
If \(g_s\) spin ladders contribute with the same frequency, each ladder carries \(\rho/g_s\), and Eq.~\eqref{eq:onsager-freq-app} gives
\begin{eqnarray}
    F = 2\pi\frac{\rho}{g_s}.
\end{eqnarray}
Thus an effectively spin-degenerate electron sequence with \(g_s=2\) gives
\(F=\pi\rho\), motivating the green guide in panels (a) and (b). The FFT alone
does not establish a unique degeneracy assignment. It instead indicates that
the \(\rho=+0.010\,\mathrm{nm}^{-2}\) trace is closer to the spin-resolved
counting scale, whereas the \(\rho=+0.020\,\mathrm{nm}^{-2}\) trace contains a
strong spin-degenerate-like component. We therefore examine the fixed-density
chemical potential and the nearest-Fermi LL spacing in
Fig.~\ref{fig:electron-mu-splitting-comparison} as a direct LL-level diagnostic
of the different electron-side spectra in
Figs.~\ref{fig:magnetization-fft}(a)--\ref{fig:magnetization-fft}(d).
Panel (a) shows that the fixed-density chemical potential is determined from the same full spin-sector LL spectrum for both \(\rho=+0.010\) and \(+0.020\,\mathrm{nm}^{-2}\).
Panel (b) then compares the nearest-Fermi-energy LL spacing \(\Delta E_{\mathrm{near}}\), defined by the two LLs immediately below and above \(\mu_{\mathcal S}\).
For \(\rho=+0.010\,\mathrm{nm}^{-2}\), the near-Fermi spacing is comparatively
regular, consistent with the dominant FFT weight near the spin-resolved
counting scale \(F=2\pi\rho\). For
\(\rho=+0.020\,\mathrm{nm}^{-2}\), \(\Delta E_{\mathrm{near}}\) has a much
stronger oscillatory modulation over the same \(1/B\) range, alternating
between small spin splittings and larger inter-LL spacings. The corresponding
chemical-potential modulation is visible in panel (a) of
Fig.~\ref{fig:electron-mu-splitting-comparison}. Together, these observations
support---without uniquely proving---the spin-degenerate-like interpretation
of the low-frequency FFT component.

On the hole side, the anomalous LLs from the two spin sectors are split into branches on opposite sides of \(E_0\), so they do not form the same kind of nearly spin-degenerate sequence that appears for the electron-side dispersive LLs.
The FFT spectra in Figs.~\ref{fig:magnetization-fft}(e)--\ref{fig:magnetization-fft}(l) therefore generally show one main counting peak.
Compared with the electron-side spectra, however, the hole-side peaks are much broader.
For \(\rho=-0.020\) and \(-0.140\,\mathrm{nm}^{-2}\), the \(\delta\Omega_\rho\) spectra in panels (g) and (i) are so broad that the peak position is difficult to assign reliably.
This is consistent with the direct \(1/B\)-domain traces in Fig.~\ref{fig:thermo-wide-osc}(g,i), where the \(\delta\Omega_\rho\) oscillations decay rapidly as \(1/B\) increases.
The corresponding magnetization spectra in Figs.~\ref{fig:magnetization-fft}(h) and \ref{fig:magnetization-fft}(j) still show visible peaks, but these peaks remain broader.
For \(\rho=-0.010\) and \(-0.150\,\mathrm{nm}^{-2}\), the main FFT peaks can be resolved in both \(\delta\Omega_\rho\) and \(\widetilde M_{\mathrm{no0LL}}\), and their positions are consistent with the active-density counting frequency \(F=2\pi\rho_{\mathcal S}\) in Eq.~\eqref{eq:onsager-freq-app}, shown by the black dashed lines.

\subsection{Local window selection}

\begin{figure}[htbp]
    \centering
    \includegraphics[width=\linewidth]{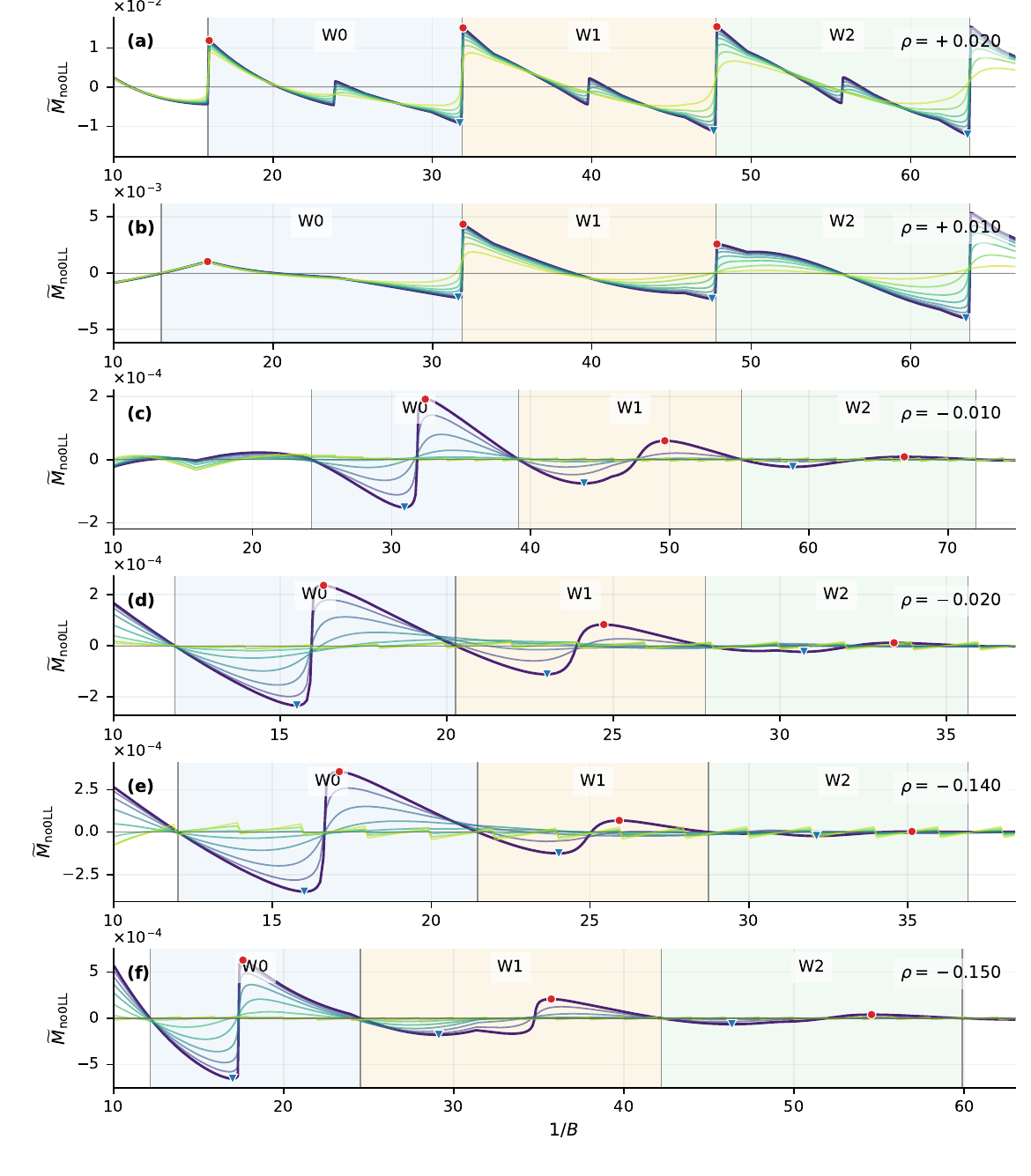}
    \caption{Local window selection for the zero-mode-excluded oscillatory magnetization at $\Lambda=1.0\,\mathrm{nm}^{-1}$. Each row shows $\widetilde M_{\mathrm{no0LL}}(1/B)$ for one density, with the same temperature sequence as Fig.~\ref{fig:thermo-wide-osc}. Shaded regions mark the three windows $W_0$, $W_1$, and $W_2$ used for local amplitude extraction; red circles and blue triangles mark the maximum and minimum of the lowest-temperature trace in each window.}
    \label{fig:local-window-selection}
\end{figure}

To resolve the thermal damping factor, we need to extract the oscillation amplitude as a function of temperature for the local $1/B$ windows of strong oscillation period, rather than a single global FFT. 
The windows are chosen from the lowest-temperature $\widetilde M_{\mathrm{no0LL}}$ trace and then held fixed for all temperatures.
Operationally, we first identify zero crossings of the oscillatory trace and use same-slope zero crossings to define full local cycles. Fig.~\ref{fig:local-window-selection} shows the resulting $W_0$, $W_1$, and $W_2$ windows.
For $\rho=+0.020\,\mathrm{nm}^{-2}$,
Fig.~\ref{fig:magnetization-fft}(b) shows dominant magnetization weight near
the lower frequency $F=\pi\rho$. The windows in
Fig.~\ref{fig:local-window-selection}(a) therefore follow the corresponding
dominant zero-to-zero cycles; the smaller shoulders are not treated as
independent windows.

Table~\ref{tab:local-window-periods} compares the selected windows with the expected period $1/F$.
For $\rho=+0.020\,\mathrm{nm}^{-2}$ we track only the main spin-degenerate-like LL oscillation in Fig.~\ref{fig:local-window-selection}(a), so the table uses \(F=\pi\rho\).
For the other densities, the windows follow the spin-resolved active-density counting frequency, \(F=2\pi\rho_{\mathcal S}\), with \(\rho_{\mathcal S}=|\rho|\) for the electron and shallow-hole rows and \(\rho_{\mathcal S}=\rho_{\mathrm{low},e}\) for the deep-hole rows.
Because the numerical waveform is not a perfect sinusoid, individual zero-aligned windows can be slightly uneven; the final column checks the total span of the three selected windows against $3/F$ and gives a value close to unity for all densities, suggesting the correct assignment of the local windows to the expected oscillation period.

\begin{table}[htbp]
    \caption{Local window ranges in $x=1/B$ for Fig.~\ref{fig:local-window-selection}. For $\rho=+0.020\,\mathrm{nm}^{-2}$, the period is evaluated using the main spin-degenerate-like frequency $F=\pi\rho$; for the other rows, it is evaluated using the spin-resolved active-density frequency $F=2\pi\rho_{\mathcal S}$. The last column compares the combined three-window span with three ideal periods.}
    \label{tab:local-window-periods}
    \centering
    \begin{ruledtabular}
    \begin{tabular}{c c c c c c}
        $\rho$ & $W_0$ & $W_1$ & $W_2$ & $1/F$ & $\sum_i\Delta x_i/(3/F)$ \\
        \hline
        $+0.020$ & $[15.933,31.862]$ & $[31.862,47.784]$ & $[47.784,63.703]$ & $15.915$ & $1.000$ \\
        $+0.010$ & $[12.992,31.857]$ & $[31.857,47.787]$ & $[47.787,63.700]$ & $15.915$ & $1.062$ \\
        $-0.010$ & $[24.238,39.150]$ & $[39.150,55.161]$ & $[55.161,71.995]$ & $15.915$ & $1.000$ \\
        $-0.020$ & $[11.842,20.265]$ & $[20.265,27.766]$ & $[27.766,35.639]$ & $7.958$ & $0.997$ \\
        $-0.140$ & $[12.036,21.459]$ & $[21.459,28.727]$ & $[28.727,36.892]$ & $8.309$ & $0.997$ \\
        $-0.150$ & $[12.157,24.523]$ & $[24.523,42.184]$ & $[42.184,59.880]$ & $17.385$ & $0.915$ \\
    \end{tabular}
    \end{ruledtabular}
\end{table}

\subsection{Thermal damping of Normalized oscillation amplitude}

\begin{figure}[htbp]
    \centering
    \includegraphics[width=\linewidth]{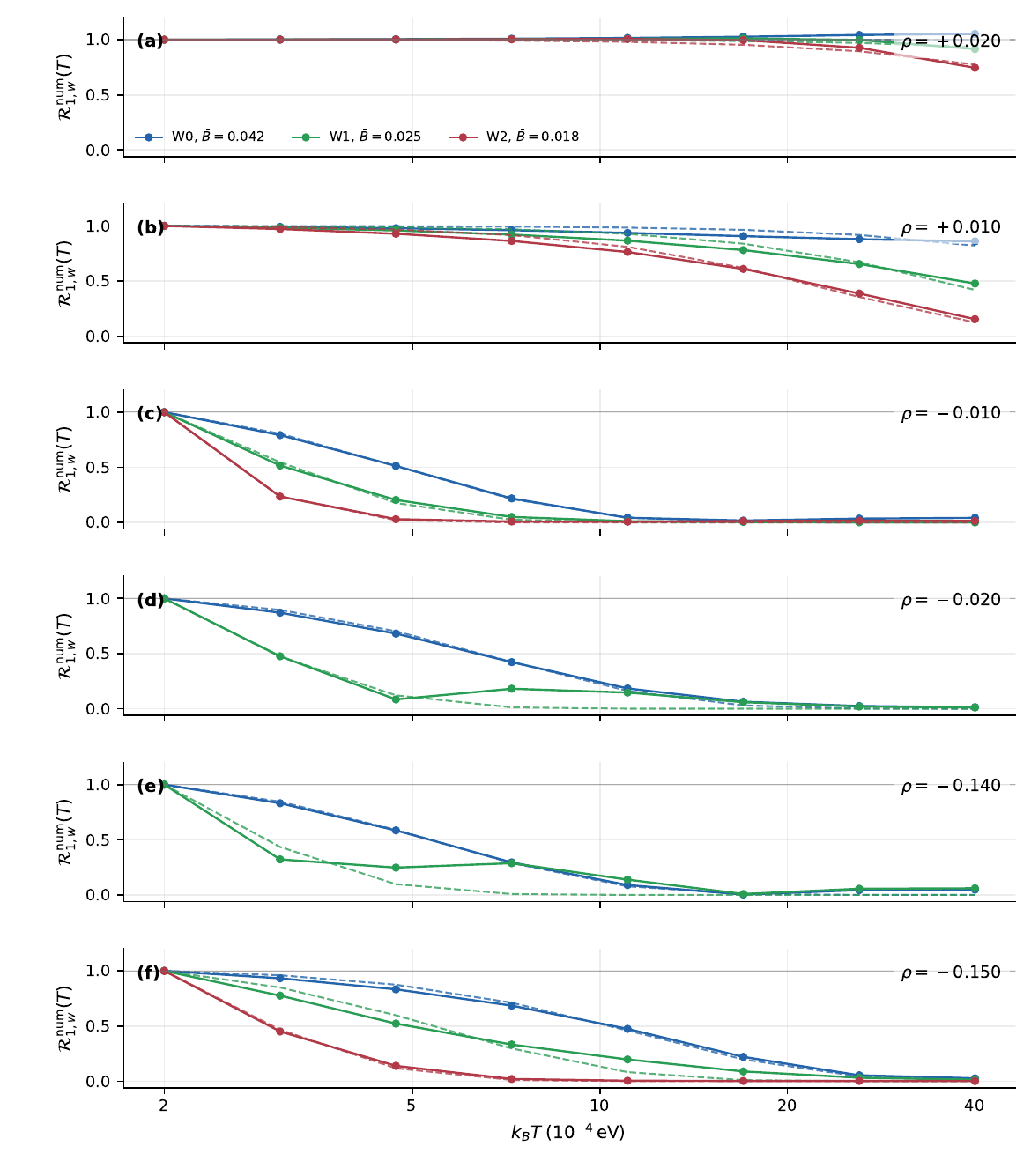}
    \caption{Windowed \(p=1\) thermal damping extracted from the Fig.~\ref{fig:local-window-selection} windows. Solid markers show the normalized numerical amplitude \(\mathcal R^{\mathrm{num}}_{1,w}(T)\) defined in Eq.~\eqref{eq:window-p1-normalized-app}; dashed curves are one-parameter standard-LK fits for the same window. Three colors blue, green and red label \(W_0\), \(W_1\), and \(W_2\), respectively, and the legend gives the representative field \(\bar B_w\) for the first row. The \(W_2\) series for \(\rho=-0.020\) and \(-0.140\,\mathrm{nm}^{-2}\) are omitted because their amplitudes are too small for reliable damping extraction.}
    \label{fig:window-p1-damping}
\end{figure}

Figure~\ref{fig:window-p1-damping} is obtained directly from the windowed oscillatory magnetization in Fig.~\ref{fig:local-window-selection}.
For each density, temperature, and local window \(w=W_0,W_1,W_2\), we first restrict the background-subtracted trace \(\widetilde M_{\mathrm{no0LL}}(x,T)\), with \(x=1/B\), to the window interval \([x_{w,0},x_{w,1}]\) listed in Table~\ref{tab:local-window-periods}.
The frequency \(F_w\) is the same active-density counting frequency used to choose the window: for \(\rho=+0.020\,\mathrm{nm}^{-2}\) it is the main spin-degenerate-like frequency \(F_w=\pi\rho\), while for the other rows it is \(F_w=2\pi\rho_{\mathcal S}\).
Within this fixed window, the fundamental \(p=1\) harmonic is extracted by the linear projection
\begin{equation}
    \widetilde M_{\mathrm{no0LL}}(x,T)
    =
    C_{1,w}(T)\cos(2\pi F_w x)+S_{1,w}(T)\sin(2\pi F_w x)+C_{0,w}(T).
    \label{eq:window-p1-fit-app}
\end{equation}
The numerical amplitude plotted by the solid markers is
\begin{equation}
    \mathcal R_{1,w}^{\mathrm{num}}(T)
    =
    \frac{A_{1,w}(T)}{A_{1,w}(T_{\min})},
    \qquad
    A_{1,w}(T)=\sqrt{C_{1,w}^{2}(T)+S_{1,w}^{2}(T)}.
    \label{eq:window-p1-normalized-app}
\end{equation}
For \(\rho=-0.020\) and \(-0.140\,\mathrm{nm}^{-2}\), the \(W_2\) amplitudes are too small for a reliable normalized damping curve, so those two \(W_2\) series are not shown.

The dashed curves in Fig.~\ref{fig:window-p1-damping} are then obtained by fitting the normalized amplitudes in each window to the standard LK thermal form
\begin{equation}
    \mathcal R_{1,w}^{\mathrm{LK}}(T;m_w)
    =
    \frac{R_T(T,m_w,\bar B_w)}
    {R_T(T_{\min},m_w,\bar B_w)},
    \qquad
    R_T=\frac{X}{\sinh X},
    \qquad
    X=\frac{2\pi^2 k_B T\,m_w}{\bar B_w}.
    \label{eq:window-lk-fit-app}
\end{equation}
Here \(\bar B_w=1/[(x_{w,0}+x_{w,1})/2]\) is the representative field of the window, and \(m_w\) is the effective mass.
Here we treat \(m_w\) as a single free fitting parameter, rather than using the branch-resolved analytical mass from Sec.~\ref{app:LL-spacing-formulas}.
The fitted \(m_w\) is summarized and compared with the branch-resolved expectation in Fig.~\ref{fig:window-meff-summary}.

The branch-resolved LK expressions explain the scales and behaviors of these fitted masses.
Using the local spacings from Sec.~\ref{app:LL-spacing-formulas}, the LK thermal factor for a single branch is
\begin{equation}
    R_T^{\alpha,\eta}(p,T)
    =
    \frac{X_p^{\alpha,\eta}}{\sinh X_p^{\alpha,\eta}},
    \qquad
    X_p^{\alpha,\eta}
    =
    \frac{2\pi^2 p\kB T}{v_\mu^{\alpha,\eta}}.
    \label{eq:RT-branch-app}
\end{equation}
Writing $\dmu=\mu-E_0$ and substituting the upper-branch spacings from Eq.~\eqref{eq:dEdn-upper-app} together with the lower-branch spacings from Eqs.~\eqref{eq:dEdn-LLminus-app} and \eqref{eq:dEdn-LLTR-app}, the corresponding branch-resolved arguments are
\begin{align}
    X_p^{\uparrow,+}
    &=
    \frac{\pi^2 p\kB T}{aB}
    \frac{\dmu^2-c^2B}{\dmu^2},
    \qquad
    X_p^{\downarrow,+}
    =
    \frac{\pi^2 p\kB T}{aB}
    \frac{\dmu^2+c^2B}{\dmu^2},
    \label{eq:Xp-normal-branch-app}
    \\
    X_p^{\uparrow,-}
    &=
    \frac{\pi^2 p\kB T}{aB}
    \frac{c^2B-\dmu^2}{\dmu^2},
    \qquad
    X_p^{\downarrow,-}
    =
    \frac{\pi^2 p\kB T}{aB}
    \frac{\dmu^2+c^2B}{\dmu^2}.
    \label{eq:Xp-anomalous-branch-app}
\end{align}
The upper-branch expressions apply in the normal dispersive regime, while the lower-branch expressions apply in the anomalous regime within their corresponding crossing windows.
For $\dmu^2\gg c^2B$, both normal expressions reduce to the parabolic result $X_p\simeq\pi^2p\kB T/(aB)$.
For anomalous LLs in the lower-branch crossing windows, \(0<\dmu<c\sqrt{B}\) for \(\ELLminus\) and \(-c\sqrt{B}<\dmu<0\) for \(\ELLTRminus\), the small local spacing makes \(X_p\) much larger at the same \(B\) and \(T\).
In the anomalous regime, where one lower branch dominates a given crossing window, the normalized damping is therefore well represented by the single-branch ratio
\begin{equation}
    \mathcal{R}^{\alpha,-}_{p=1}(\rho,B,T)
    =
    \frac{R_T^{\alpha,-}(p=1,\rho,B,T)}
    {R_T^{\alpha,-}(p=1,\rho,B,T_{\min})}.
    \label{eq:RT-normalized-app-anomalousLL}
\end{equation}
For normal electron doping, the two upper branches lie in the same energy range and are counted together at fixed density.
The fixed-density chemical potential is obtained from
\begin{equation}
    \rho
    =
    \frac{B}{2\pi}
    \sum_{\alpha=\uparrow,\downarrow}
    \sum_{n}
    f\!\left(E^\alpha_{+}(n,B)-\mu_{\rho}\right),
    \label{eq:normal-fixed-density-count-app}
\end{equation}
and the two upper-branch contributions combine at the harmonic-amplitude level as
\begin{equation}
    A^{+}_{p}(\rho,B,T)
    \propto
    \sum_{\alpha=\uparrow,\downarrow}
    v_\mu^{\alpha,+}
    R_T^{\alpha,+}(p,T)
    \cos\!\left(2\pi p n^*_{\alpha,+}\right),
    \qquad
    \mathcal{R}_p^{+}(\rho,B,T)
    =
    \frac{\left|A^{+}_{p}(\rho,B,T)\right|}
    {\left|A^{+}_{p}(\rho,B,T_{\min})\right|}.
    \label{eq:normal-combined-RT-app}
\end{equation}
Here \(E^\alpha_{+}(n^*_{\alpha,+},B)=\mu_{\rho}\).
Equation~\eqref{eq:normal-combined-RT-app} gives the coherent physical harmonic expected from the analytical LK expression, whereas Fig.~\ref{fig:window-p1-damping} uses the simpler fitted form in Eq.~\eqref{eq:window-lk-fit-app} as a local numerical diagnostic of the thermal scale.
The normal electron densities show weak thermal damping over this temperature range, whereas the anomalous-hole densities lose their \(p=1\) amplitude much more rapidly.

\subsection{Effective masses from the windowed damping}

\begin{figure}[htbp]
    \centering
    \includegraphics[width=\linewidth]{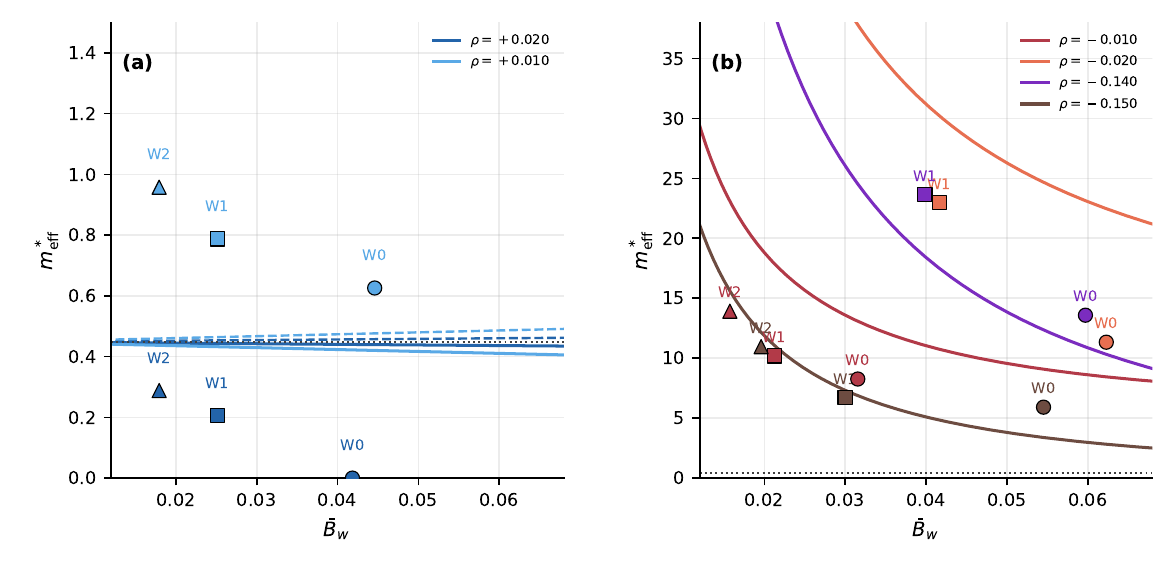}
    \caption{Effective masses extracted from the Fig.~\ref{fig:window-p1-damping} standard-LK fits. Panel (a) shows the normal electron windows for \(\rho=+0.020\) and \(+0.010\,\mathrm{nm}^{-2}\), with analytical curves for the nearby \(\ELLplus\) and \(\ELLTRplus\) upper branches; the near-zero \(W_0\) point for \(\rho=+0.020\,\mathrm{nm}^{-2}\) reflects the nearly non-damping projected \(p=1\) amplitude in that window. Panel (b) shows the anomalous-hole windows for \(\rho=-0.010,-0.020,-0.140\), and \(-0.150\,\mathrm{nm}^{-2}\), with analytical curves for the active lower branch selected by the fixed-density counting; the \(W_2\) points for \(\rho=-0.020\) and \(-0.140\,\mathrm{nm}^{-2}\) are omitted because their amplitudes are too small for reliable extraction. Markers are fitted masses for the local windows, labelled by \(W_0,W_1,W_2\) when shown. The dotted horizontal line marks the parabolic value \(1/(2a)\).}
    \label{fig:window-meff-summary}
\end{figure}

The fitted standard-LK masses from the dashed curves in Fig.~\ref{fig:window-p1-damping} are summarized in Table~\ref{tab:window-meff-fit-summary} and Fig.~\ref{fig:window-meff-summary}. The standard-LK fits in Fig.~\ref{fig:window-p1-damping} use only the local \(p=1\) harmonic and should therefore be read as qualitative thermal-scale diagnostics rather than precision measurements of a unique band mass. Because the numerical waveform is not a pure sinusoid, as clearly shown in Fig.~\ref{fig:local-window-selection}, the higher harmonics \(p>1\) is also not negligible, but we only focus on $p=1$ harmonic in the current analysis of effective mass below. 

Each marker in Fig.~\ref{fig:window-meff-summary} corresponds to the extracted effective mass for one window and is placed at the representative field \(\bar B_w\).
The curves show the branch-resolved analytical expectation \(m^*_{\mathrm{eff}}(B)=B/v_\mu\) from Sec.~\ref{app:lk-branch}; for normal electron densities we show both nearby upper branches, while for anomalous-hole densities we show the active lower branch selected by the fixed-density counting.
Figure~\ref{fig:window-meff-summary}(a) shows the effective mass for the normal LLs in electron carrier regime.
Except for the first \(\rho=+0.020\,\mathrm{nm}^{-2}\) window, the fitted masses remain of order unity and lie around the parabolic scale \(1/(2a)\), with visible window-to-window variation because the two upper branches, \(\ELLplus\) and \(\ELLTRplus\), contribute in the same energy range and are not exactly degenerate. For \(\rho=+0.020\,\mathrm{nm}^{-2}\), we note that the projected \(p=1\) amplitude is nearly temperature independent and even increases slightly with temperature, as shown in Fig.~\ref{fig:window-p1-damping}a, so the one-parameter LK fit returns a mass close to zero. This suggests the higher harmonics ($p>1$) also plays important role in the temperature dependence of the oscillation amplitude in this window, beyond the current $p=1$ harmonics fitting. Fig.~\ref{fig:window-meff-summary}(b) shows the effective mass of anomalous LLs in hole carrier regime.
The fitted masses are much larger and more field dependent, qualitatively following the trend of the active lower-branch analytical curves. This is the mass-scale version of the rapid thermal damping in Fig.~\ref{fig:window-p1-damping}: the anomalous LL spacing is much smaller, so the LK argument \(X\propto m^*_{\mathrm{eff}}T/B\) is much larger at comparable \(B\) and \(T\). However, the fitted masses are quantitatively deviated from the analytical branch-resolved curves in Fig.~\ref{fig:window-meff-summary}(b), suggesting the limitation of the effective mass description of the thermal damping in the anomalous LL regime.

\begin{table}[htbp]
    \caption{Fitted standard-LK masses \(m_w\) extracted from the dashed curves in Fig.~\ref{fig:window-p1-damping} and plotted in Fig.~\ref{fig:window-meff-summary}. The entries correspond to the local windows \(W_0,W_1,W_2\) defined in Fig.~\ref{fig:local-window-selection}. }
    \label{tab:window-meff-fit-summary}
    \begin{ruledtabular}
    \begin{tabular}{ccccc}
        \(\rho\,(\mathrm{nm}^{-2})\) & regime & \(m_{W_0}\) & \(m_{W_1}\) & \(m_{W_2}\) \\
        \hline
        \(+0.020\) & normal LL & \(1.04\times10^{-4}\) & \(0.206\) & \(0.288\) \\
        \(+0.010\) & normal LL & \(0.626\) & \(0.788\) & \(0.958\) \\
        \(-0.010\) & anomalous LL & \(8.27\) & \(10.2\) & \(13.9\) \\
        \(-0.020\) & anomalous LL & \(11.3\) & \(23.0\) & -- \\
        \(-0.140\) & anomalous LL & \(13.6\) & \(23.7\) & -- \\
        \(-0.150\) & anomalous LL & \(5.92\) & \(6.73\) & \(11.0\) \\
    \end{tabular}
    \end{ruledtabular}
\end{table}

\bibliography{refs}

\end{document}